\documentclass{aastex62}

\received{\today}
\revised{\today}
\accepted{\today}

\submitjournal{AJ}

\shorttitle{Eclipsing Binaries in the Northern Galactic Plane}
\shortauthors{F.-Z. Ren et al.}

\begin{document}

\title{Eclipsing Binary Populations across the Northern Galactic Plane from the KISOGP survey}

\correspondingauthor{Fangzhou Ren, Huawei Zhang, and Richard de Grijs}
\email{renfz@pku.edu.cn \\ zhanghw@pku.edu.cn \\ richard.de-grijs@mq.edu.au}

\author{Fangzhou Ren}
\affiliation{Department of Astronomy, School of Physics, Peking
  University, Yi He Yuan Lu 5, Hai Dian District, Beijing 100871,
  People's Republic of China}
\affiliation{Kavli Institute for Astronomy and Astrophysics, Peking
  University, Yi He Yuan Lu 5, Hai Dian District, Beijing 100871,
  People's Republic of China}

\author[0000-0002-7203-5996]{Richard de Grijs}
\affiliation{Department of Physics and Astronomy, Macquarie
  University, Balaclava Road, Sydney, NSW 2109, Australia}
\affiliation{Research Centre for Astronomy, Astrophysics and
  Astrophotonics, Macquarie University, Balaclava Road, Sydney, NSW
  2109, Australia}
\affiliation{International Space Science Institute--Beijing, 1
  Nanertiao, Zhongguancun, Hai Dian District, Beijing 100190, People's
  Republic of China}

\author[0000-0002-7727-1699]{Huawei Zhang}
\affiliation{Department of Astronomy, School of Physics, Peking
  University, Yi He Yuan Lu 5, Hai Dian District, Beijing 100871,
  People's Republic of China}
\affiliation{Kavli Institute for Astronomy and Astrophysics, Peking
  University, Yi He Yuan Lu 5, Hai Dian District, Beijing 100871,
  People's Republic of China}

\author[0000-0001-9073-9914]{Licai Deng}
\affiliation{CAS Key Laboratory of Optical Astronomy, National
  Astronomical Observatories, Chinese Academy of Sciences, Beijing
  100101, People's Republic of China}

\author[0000-0001-7084-0484]{Xiaodian Chen}
\affiliation{CAS Key Laboratory of Optical Astronomy, National
  Astronomical Observatories, Chinese Academy of Sciences, Beijing
  100101, People's Republic of China}

\author{Noriyuki Matsunaga}
\affiliation{Department of Astronomy, The University of Tokyo, 7-3-1
  Hongo, Bunkyo-ku, Tokyo 113-0033, Japan}
\affiliation{Laboratory of Infrared High-resolution Spectroscopy
  (LiH), Koyama Astronomical Observatory, Kyoto Sangyo University,
  Motoyama, Kamigamo, Kita-ku, Kyoto 603-8555, Japan}

\author[0000-0002-1802-6917]{Chao Liu}
\affiliation{CAS Key Laboratory of Optical Astronomy, National
  Astronomical Observatories, Chinese Academy of Sciences, Beijing
  100101, People's Republic of China}

\author[0000-0002-3279-0233]{Weijia Sun}
\affiliation{Department of Astronomy, School of Physics, Peking
  University, Yi He Yuan Lu 5, Hai Dian District, Beijing 100871,
  People's Republic of China}
\affiliation{Kavli Institute for Astronomy and Astrophysics, Peking
  University, Yi He Yuan Lu 5, Hai Dian District, Beijing 100871,
  People's Republic of China}
\affiliation{Department of Physics and Astronomy, Macquarie
  University, Balaclava Road, Sydney, NSW 2109, Australia}
\affiliation{Research Centre for Astronomy, Astrophysics and
  Astrophotonics, Macquarie University, Balaclava Road, Sydney, NSW
  2109, Australia}

\author[0000-0003-0332-0811]{Hiroyuki Maehara}
\affiliation{Okayama Branch Office, Subaru Telescope, National
  Astronomical Observatory of Japan, NINS, Kamogata, Asakuchi,
  Okayama, Japan}
\affiliation{Okayama Observatory, Kyoto University, 3037-5 Honjo,
  Kamogata, Asakuchi, Okayama 719-0232, Japan}

\author{Nobuharu Ukita}
\affiliation{Okayama Astrophysical Observatory, National Astronomical
  Observatory of Japan, 3037-5 Honjo, Kamogata, Asakuchi, Okayama
  719-0232, Japan}

\author{Naoto Kobayashi}
\affiliation{Laboratory of Infrared High-resolution Spectroscopy
  (LiH), Koyama Astronomical Observatory, Kyoto Sangyo University,
  Motoyama, Kamigamo, Kita-ku, Kyoto 603-8555, Japan}
\affiliation{Kiso Observatory, Institute of Astronomy, School of
  Science, The University of Tokyo, 10762-30 Mitake, Kiso-machi,
  Kiso-gun, Nagano 397-0101, Japan}
\affiliation{Institute of Astronomy, Graduate School of Science, The
  University of Tokyo, 2-21-1 Osawa, Mitaka, Tokyo 181-0015, Japan}

\begin{abstract}
We present a catalog of eclipsing binaries in the northern Galactic
Plane from the Kiso Wide-Field Camera Intensive Survey of the Galactic
Plane (KISOGP). We visually identified 7055 eclipsing binaries spread
across $\sim$330 square degrees, including 4197 W Ursa Majoris/EW-,
1458 $\beta$ Lyrae/EB-, and 1400 Algol/EA-type eclipsing binaries. For
all systems, $I$-band light curves were used to obtain accurate system
parameters. We derived the distances and extinction values for the
EW-type objects from their period--luminosity relation. We also
obtained the structure of the thin disk from the distribution of our
sample of eclipsing binary systems, combined with those of high-mass
star-forming regions and Cepheid tracers. We found that the thin disk
is inhomogeneous in number density as a function of Galactic
longitude. Using this new set of distance tracers, we constrain the
detailed structure of the thin disk. Finally, we report a global
parallax zero-point offset of $ \Delta \pi=-42.1\pm1.9\mbox{
(stat.)}\pm12.9\mbox{(syst.)}$ $\mu$as between our
carefully calibrated EW-type eclipsing binary positions and those
provided by {\sl Gaia} Early Data Release 3.  Implementation of the
officially recommended parallax zero-point correction results in a
significantly reduced offset. Additionally, we provide a photometric
characterization of our EW-type eclipsing binaries that can be applied
to further analyses.
\end{abstract}


\section{Introduction} \label{sec:intro}

Eclipsing binary systems (EBS) exhibit optical variability because of
geometric properties rather than intrinsic physical changes. EBS
encompass almost all stages of binary evolution, covering timescales
as long as 1--10 Gyr. This explains their large numbers in the
Galaxy. EBS analysis offers a good opportunity to obtain precise
fundamental physical parameters from their system properties by means
of photometric and/or spectroscopic observations---including their
periods and distances, as well as accurate parameters for their
components, like masses and radii
\citep[e.g.][]{2010A&ARv..18...67T}. This enables us to study unique
aspects of their stellar evolution and stellar activity.

EBS can be divided into three types based on their light curve shapes,
i.e., Algol (EA)-, $\beta$ Lyrae (EB)-, and W Ursa Majoris (EW)-type
EBS. The total luminosity of EA-type EBS remains almost constant
outside the eclipses. EB types exhibit a continuous change in their
total brightness outside eclipses while the depth of the secondary
minimum is usually considerably smaller than that of the primary
minimum. Meanwhile, EW-type light curves are characterized by a smooth
shape with symmetric eclipses, and some evolved systems show
sinusoidal-like shapes.

EW-type EBS can also be used as reliable distance indicators within
the Milky Way. Since their two components fill the system's Roche
lobes, their overall visual magnitude is related to the system's
orbital period (based on Roche lobe theory), which leads to a
well-defined period--luminosity relation
(PLR). \citet{1994PASP..106..462R} derived the first calibration of
such a PLR based on 18 systems, which they eventually improved to an
accuracy of 12\% \citep{1997PASP..109.1340R}. Recently,
\citet{2018ApJ...859..140C} established PLRs in 12 optical to
mid-infrared passbands based on Tycho--{\sl Gaia} astrometric solution
(TGAS) parallaxes, reaching an improved accuracy of 8\%. PLRs provide
us with a means to determine the distances to EW-type EBS using only
photometric light curves. Therefore, as one of the most numerous types
of variable systems, EW-type EBS could be used as important Galactic
distance tracers \citep{2018SSRv..214...74M}.

Observations of EBS have a long history. Many ancient cultures
observed eclipsing systems \citep[e.g.][]{2013ApJ...773....1J}.  In
modern astronomy, early-20th century measurements of both EBS and
other variables were usually reported in papers discussing individual
objects. After the 1980s, large surveys commenced, including the
MAssive Compact Halo Objects (MACHO) survey
\citep{1995ASPC...83..221C} and the Optical Gravitational Lensing
Experiment
\citep[OGLE;][]{2011AcA....61..103G,2013AcA....63..323P,2013AcA....63..115P,2016AcA....66..421P,2016AcA....66..405S}.
The number of known EBS and other types of variable stars hence
experienced a period of explosive growth. With the development of
wide-field cameras, surveys that constantly monitor the entire
accessible sky photometrically became commonplace, e.g. the All Sky
Automated Survey
\citep[ASAS;][]{1997AcA....47..467P,2006MNRAS.368.1311P} and the
Robotic Optical Transient Search Experiment
\citep[ROTSE;][]{2000AJ....119.1901A}, part of the Northern Sky
Variability Survey
\citep[NSVS;][]{2000AJ....119.1901A,2004AJ....128.2965W,2008AJ....136.1067H,2009AJ....138..466H}.
In the near-infrared (NIR), the VISTA Variables in the V\'{i}a
L\'{a}ctea (VVV) Survey offers a less severely reddened window into
EBS projected toward the Galactic Center
\citep[e.g.][]{2010NewA...15..433M,2015AJ....149...99A}. In addition,
some all-sky surveys, such as near-Earth object (NEO) surveys whose
primary goal is to detect near-Earth asteroids and comets, are also
well suited to discover and characterize variable objects, including
EBS. This way, candidates from the Lincoln Near-Earth Asteroid
Research \citep[LINEAR;][]{2013AJ....146..101P}, the Catalina Sky
Surveys \citep[CSS;][]{2014ApJS..213....9D,2017MNRAS.469.3688D}, the
Asteroid Terrestrial-impact Last Alert System
\citep[ATLAS;][]{2018AJ....156..241H}, the All-Sky Automated Survey
for Supernovae
\citep[ASAS-SN;][]{2017PASP..129j4502K,2018MNRAS.477.3145J}, and the
Wide-field Infrared Survey Explorer
\citep[WISE;][]{2018ApJS..237...28C} have been identified.

Although the accumulation of EBS has multiplied, many previous surveys
have avoided targeting the Galactic plane due to the high extinction
there compared with less obscured regions. The few available surveys
of the Galactic plane tend to cover only a few degrees in Galactic
latitude \citep[e.g.][]{2012AN....333..706H, 2014Msngr.155...29H}.
Measurements of EBS are still lacking in the Galactic disk,
particularly in the Galactic Anti-center direction. However,
understanding the formation and evolution of the Galactic disk, as
well as its structure, is a key open issue, since in principle we can
investigate our own Galaxy's structure in much greater detail than
that of any other galaxy, thus providing a benchmark for understanding
external galaxies.

In this paper, we have collected the light curves of eclipsing binary
candidates from the Kiso Wide-Field Camera (KWFC) Intensive Survey of
the Galactic Plane (KISOGP), specifically in the northern Galactic
Plane, and classified 7055 EBSs. In Section \ref{sec:data}, we discuss
our data reduction procedures, the classification pipeline, and
several parameter distributions, e.g. of periods, magnitudes, and
eclipse depths. We also discuss the quality of our EBS sample's light
curves. Section \ref{sec:EWs} focuses on the EW types, outlining our
analysis procedure and providing a discussion of their distances and
extinction values derived from the PLR, absolute parameters, and the
structure of the thin disk mapped by EW-type systems and other tracers
(e.g. star-forming regions, Cepheids, etc.). We also discuss the
parallax zero-point offset affecting {\sl Gaia} Early Data Release 3
(eDR3) parallaxes derived from our EW-type analysis. Section
\ref{sec:conclusion} concludes the paper.

\section{Data Reduction and Results} \label{sec:data}

\subsection{Sample Selection}\label{sec:sample}

The catalog was compiled from observations taken as part of the KISOGP
survey \citep{2017EPJWC.15201027M}, acquired with the KWFC on the Kiso
105 cm Schmidt telescope at Kiso Observatory, Japan. The KWFC has been
designed for wide-field observations, taking advantage of the Kiso
Schmidt telescope's large focal-plane area. It mosaics eight CCD chips
with 8k$\times$8k pixels in total and covers an area of
2.2$\arcdeg$$\times$2.2$\arcdeg$. Typically, for any object, more than
200 exposures were acquired in the $I$ band to a depth of $I=9.5$--17
mag with a signal-to-noise ratio (S/N) better than 30. The main goal
of the survey is to study the Galactic disk using variable stars as
tracers \citep[e.g. Miras;][]{2017ApJS..232...16Y}. The KISOGP survey
covers the northern Galactic disk, which spans from $60\arcdeg$ to
$210\arcdeg$ in Galactic longitude, and from $-1\arcdeg$ to 1$\arcdeg$
in Galactic latitude. The region $70\arcdeg<l<80\arcdeg,
1\arcdeg<b<3\arcdeg$ is also covered. \citep[For the relevant maps and
  pointings, see Figure 1 of][]{2017EPJWC.15201027M}.

For each epoch, a 5 s exposure and three 60 s exposures are taken to
optimally cover the full magnitude range. The 5 s exposure is suitable
for stars at the bright end ($I \sim 9$ mag), while the 60 s exposures
are better for the faint end, from $I=11$ mag down to $I=17$ mag. A
reference target list was established by combining all
epochs. Candidate variable stars were selected by considering three
variability indices, including interquartile ranges, weighted standard
deviations, and von Neumann ratios \citep{2017MNRAS.464..274S}. Among
$\sim$8 million stars whose time-series data were examined, more than
50,000 objects were selected as candidate variable stars (N. Matsunaga
et al., in prep.). In this paper, 7752 candidate variables are
considered for our EBS selection, based on Fourier analysis, with
periods determined in Section \ref{sec:classfication}
\citep{2020ApJS..249...18C}.

\subsection{Eclipsing Binary Selection}\label{sec:classfication}

To properly identify EBS, high-quality folded light curves are
needed. The basic idea is to determine optimally constrained periods
of variability. We adopted the Lomb--Scargle (LS) periodogram
\citep{1976Ap&SS..39..447L,1982ApJ...263..835S} and String--Length
(SL) methods \citep{1965ApJS...11..216L,2002A&A...386..763C} to obtain
optimal periods for all candidates.

The input periods ranged from 0.08 to 10 days in steps of 0.00001 days
(for the LS method). The short-period limit for known EW types is 0.16
days \citep{2015AcA....65...39S}. Ultra-short-period EW types, which
usually contain highly evolved stars, usually exhibit sinusoidal-like
light curves. In this situation, the primary and secondary minima may
have similar depths. They are often misidentified as the same feature
in phased light curves. This will result in an output harmonic or
aliased period of the true period, where $P_{\rm output}$/${P_{\rm
true}}$ is a ratio of integers, usually 1:2. For this reason, we
start at one-half of 0.16 days. Any harmonic period found will be
corrected during our subsequent visual determination by multiplying
the period to make sure that two eclipses occur within a single
period.

To further improve the accuracy of our period determination, the SL
method was applied to the output periods from the LS method. We tried
2000 periods within 0.0002 days around the output period. This led to
an improvement in the fifth decimal place: the majority of our EBS
(64.1\%) have periods with errors smaller than $1 \times 10^{-5}$
days. The output periods will be used as input periods for our visual
verification.

\subsection{Visual Verification}\label{sec:deter}

\begin{figure}[htbp]
\centering
\begin{minipage}[t]{0.32\textwidth}
\centering
\includegraphics[width=1\textwidth]{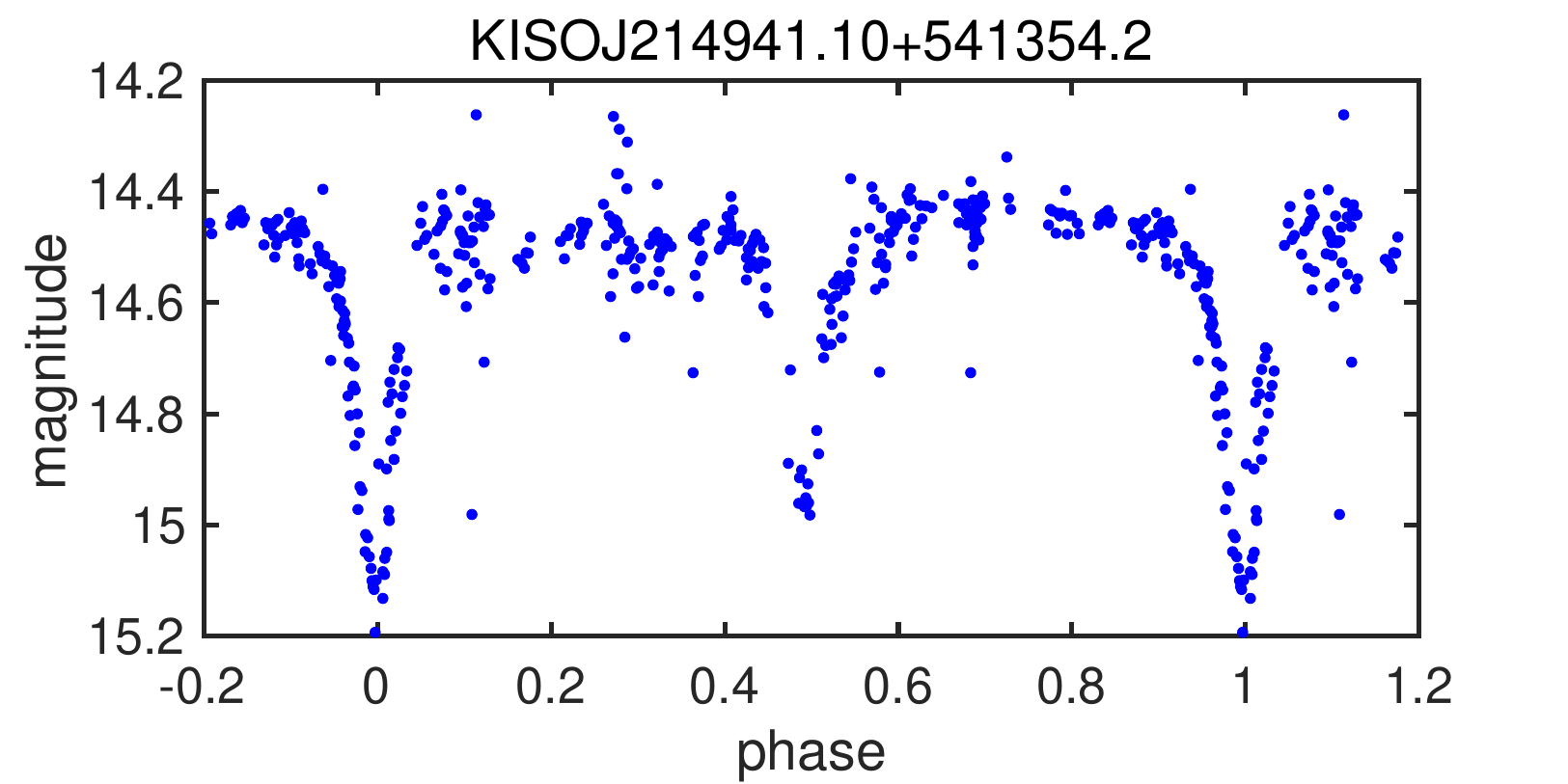}
\end{minipage}
\begin{minipage}[t]{0.32\textwidth}
\centering
\includegraphics[width=1\textwidth]{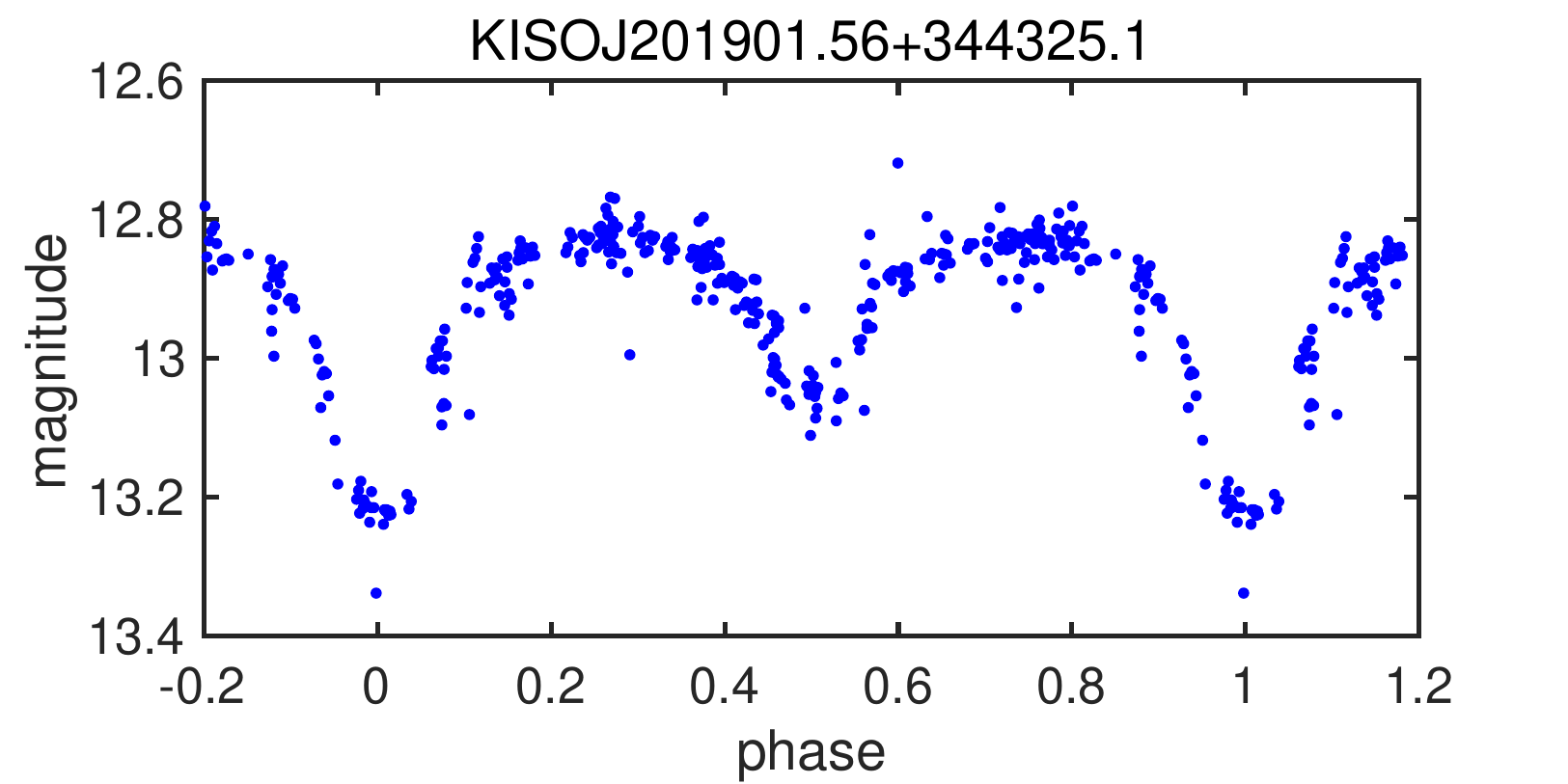}
\end{minipage}
\begin{minipage}[t]{0.32\textwidth}
\centering
\includegraphics[width=1\textwidth]{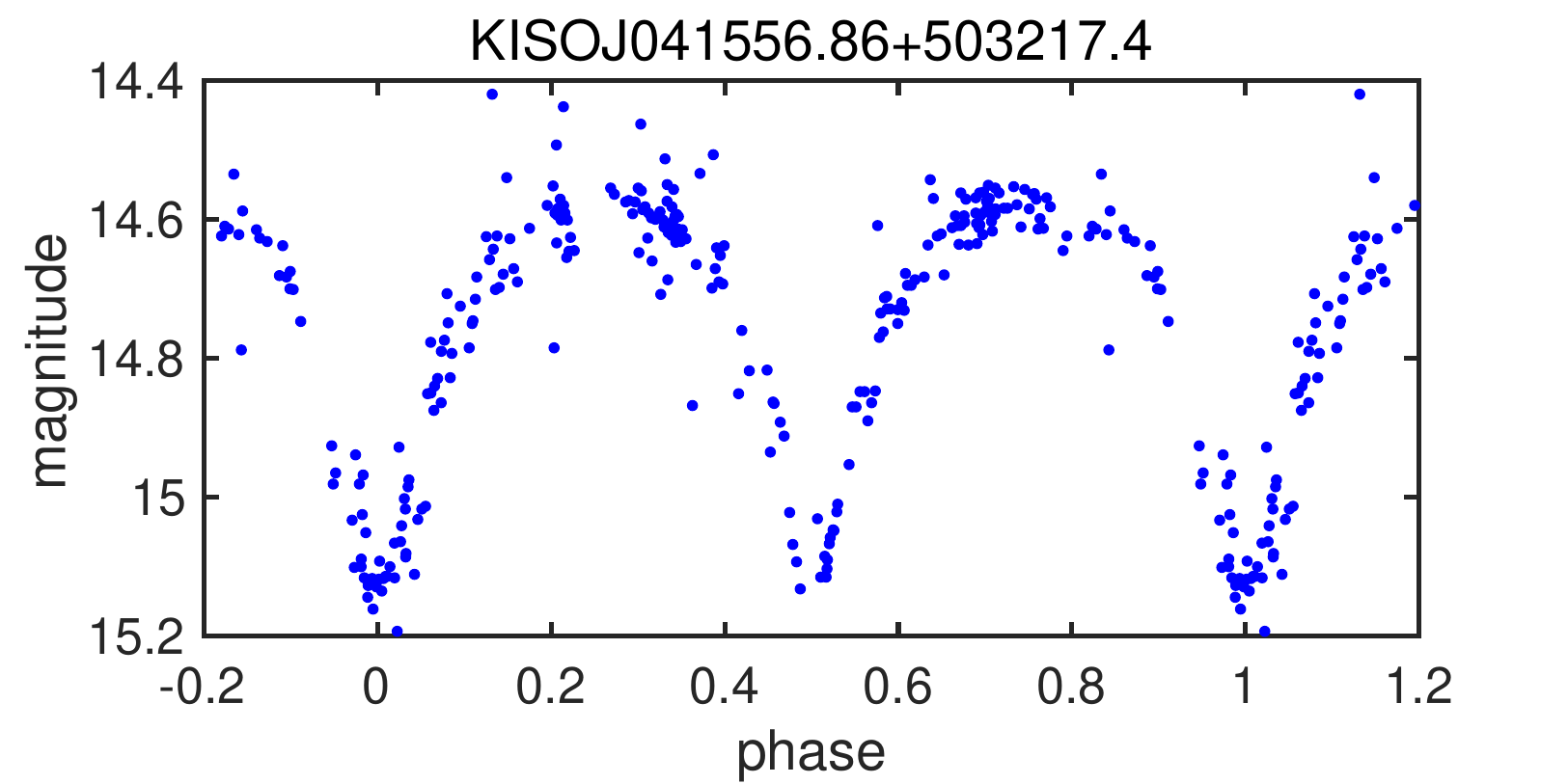}
\end{minipage}
\caption{Examples of EA-, EB-, and EW-type EBS (left to
  right). \label{fig:1}}
\end{figure}

Armed with the periods thus determined, a visual check was done of the
photometric parameters and type determination based on phased light
curves. EBS candidates were selected based on visual examination of
their light-curve morphology. We removed all objects classified as
other types of variable stars, such as RR Lyrae, Cepheids,
non-periodic variables, etc. Meanwhile, a visual check of the period
(particularly for those objects for which we need to multiply the
period to encompass two eclipses within a single period), a visual
determination of the phase of the primary eclipse (which will be
shifted to 0 and represents $t_0$), the system's magnitude outside the
eclipse, and the primary and secondary eclipses were
done. Subsequently, the objects' types were determined.

The order in which we examined the candidates was random, and all
objects were processed randomly three times. If a system's type
determination was the same and the standard errors of the parameters
were less than 0.02 mag or 0.02 in phase, the corresponding type and
the mean values of the thrice-determined parameters were adopted. For
those objects that failed to pass this round, this procedure was
repeated.

Our final catalog contains 7055 carefully visually selected EBS,
composed of 4197 EW-, 1458 EB-, and 1400 EA-type systems. Figure
\ref{fig:1} shows three randomly selected sample light curves.

\begin{deluxetable*}{ccccccccccccc}
\small
\tablecaption{Eclipsing Binaries Catalog \label{tab:1}}
\tablewidth{700pt}
\tabletypesize{\scriptsize}
\tablehead{
\colhead{ID} &
\colhead{Exposure}&
\colhead{Type} &
\colhead{Period} &
\colhead{$\rm err_{\it P}$} &
\colhead{$t_0$} &
\colhead{$I_{\rm max}$} &
\colhead{$I_{\rm pri}$} &
\colhead{$I_{\rm sec}$} &
\colhead{$\rm Ratio$}  &
\colhead{$\rm Distance$}  &
\colhead{$A_V$} &
\colhead{$\Delta m_{\rm 2MASS}$}
\\
\colhead{} &
\colhead{} &
\colhead{} &
\colhead{days} &
\colhead{$\rm 10^{-7}days$} &
\colhead{days} &
\colhead{mag} &
\colhead{mag} &
\colhead{mag} &
\colhead{}&
\colhead{kpc} & 
\colhead{mag} & 
\colhead{mag} }
\startdata
KISOJ000009.94+612905.2& 264& EA& 1.4418015& 67& 56167.64799& 13.660& 14.039& 13.833& 0.456&...& ...& ...\\
KISOJ000010.73+621848.8& 189& EW& 0.3890992& 38& 56167.69297& 15.414& 15.562& 15.538& 0.838&2.387& 0.860& 0.021\\
KISOJ000017.82+625339.6& 184& EB& 0.5820739& 23& 56167.31869& 15.552& 15.960& 15.754& 0.495&...& ...& ...\\
KISOJ000026.19+632139.0& 43& EW& 0.3180290& 393& 56175.60476& 16.333& 16.856& 16.804& 0.901&2.298& 1.469& 0.422\\
KISOJ000039.63+622214.1& 174& EW& 0.2859520& 1& 56167.59347& 16.137& 16.458& 16.390& 0.788&1.646& 1.778& 0.253\\
KISOJ000048.17+614603.1& 178& EW& 0.3024461& 1& 56167.77640& 16.016& 16.732& 16.637& 0.867&1.352& 2.612& 0.705\\
KISOJ000053.24+613059.8& 256& EW& 0.2826091& 1& 56167.60270& 16.093& 16.441& 16.378& 0.819&1.747& 1.497& 0.041\\
KISOJ000056.84+625228.4& 267& EW& 0.5717867& 12& 56167.37203& 14.228& 14.445& 14.440& 0.977&1.986& 1.657& 0.158\\
KISOJ000111.08+625139.7& 113& EB& 1.0133044& 1& 56167.36126& 16.160& 16.758& 16.492& 0.555&...& ...& ...\\
KISOJ000112.76+623859.7& 296& EA& 1.3701021& 99& 56167.30771& 14.589& 14.835& 14.819& 0.935&...& ...& ...\\
KISOJ000123.66+613746.8& 230& EW& 0.3871284& 8& 56167.49329& 15.738& 16.049& 16.004& 0.855&3.088& 0.401& 0.153\\
KISOJ000132.54+615234.2& 78& EW& 0.3660444& 36& 56251.31592& 16.772& 17.107& 17.092& 0.955&3.148& 1.900& 0.177\\
KISOJ000135.90+624456.3& 171& EB& 0.7391682& 1& 56167.17345& 16.514& 16.946& 16.700& 0.431&...& ...& ...\\
KISOJ000138.55+615348.4& 186& EA& 0.7856778& 147& 56167.74912& 12.237& 12.485& 12.405& 0.677&...& ...& ...\\
KISOJ000145.92+614214.0& 265& EB& 0.3333018& 2& 56167.67988& 12.087& 12.746& 12.382& 0.448&...& ...& ...\\
KISOJ000208.16+630633.5& 311& EW& 0.3092946& 1& 56167.58121& 12.957& 13.164& 13.141& 0.889&0.577& 0.639& 0.080\\
KISOJ000210.22+611924.2& 175& EW& 0.4997257& 21& 56167.51962& 16.190& 16.413& 16.371& 0.812&3.863& 1.878& 0.013\\
KISOJ000233.43+614514.3& 167& EW& 0.3152875& 1& 56167.71510& 16.774& 17.297& 17.199& 0.813&3.119& 1.019& 0.060\\
KISOJ000240.13+622844.5& 285& EW& 0.4282596& 3& 56167.75852& 14.515& 14.622& 14.618& 0.963&1.645& 1.237& 0.094\\
KISOJ000250.76+624850.0& 169& EW& 0.2579880& 407& 56175.50704& 15.813& 16.165& 16.162& 0.991&1.330& 1.441& 0.060\\
 \nodata & \nodata & \nodata & \nodata  & \nodata & \nodata & \nodata & \nodata & \nodata &    \nodata & \nodata & \nodata \\\enddata
\tablecomments{This table is available in its entirety in machine-readable form.}
\end{deluxetable*}

Table \ref{tab:1} includes all EBS parameters, including the system
ID, effective number of exposures, coordinates, type, period and
period error, $t_0$ in units of the modified heliocentric Julian date
(HJD-2400000.5; which sets the primary eclipse at phase zero), the
maximum magnitude outside eclipses, the magnitudes of both eclipses
within one period in the $I$ band, and the depth ratio of the two
eclipses. The objects' distances, extinction values in the $V$ band
for EW-type systems, and the magnitude differences in the Two Micron
All Sky Survey \citep[2MASS;][]{2006AJ....131.1163S} photometric
system will be described in Section \ref{sec:extinction}.

\subsection{Distribution Map}\label{sec:distribution}

\begin{figure}[htbp]
\centering
\begin{minipage}[t]{1\textwidth}
\centering
\includegraphics[width=0.95\textwidth]{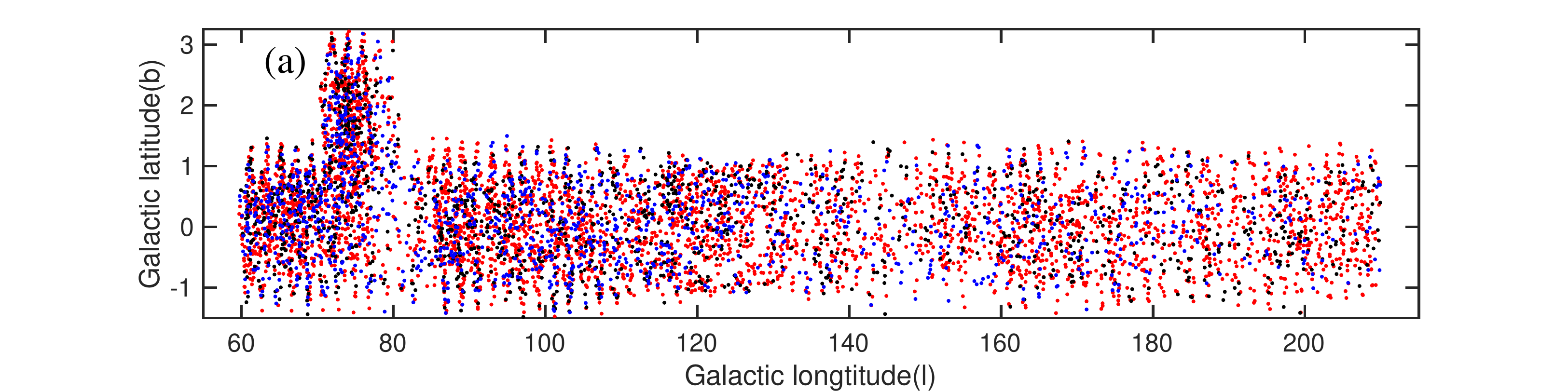}
\end{minipage}

\begin{minipage}[t]{0.32\textwidth}
\centering
\includegraphics[width=1\textwidth]{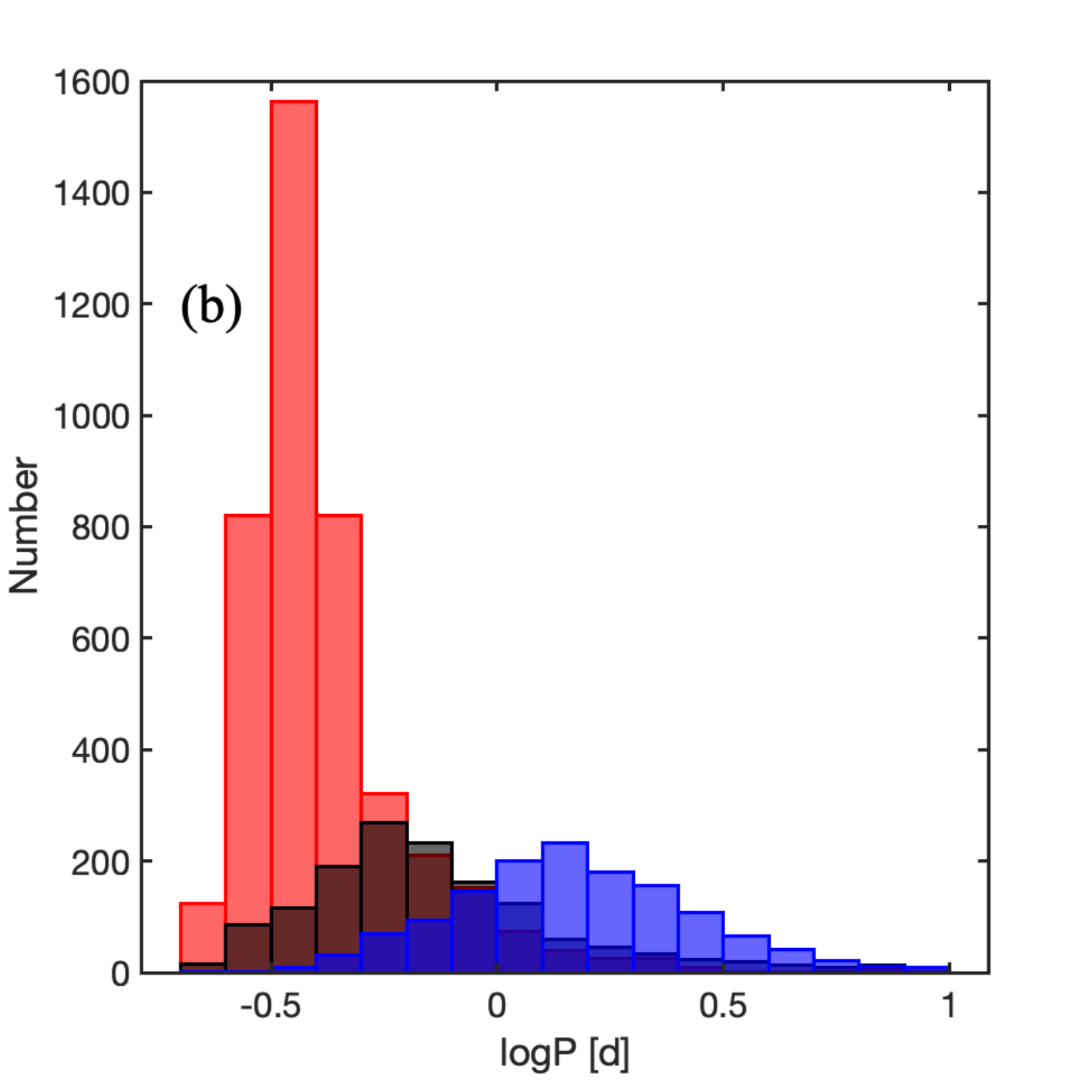}
\end{minipage}
\begin{minipage}[t]{0.32\textwidth}
\centering
\includegraphics[width=1\textwidth]{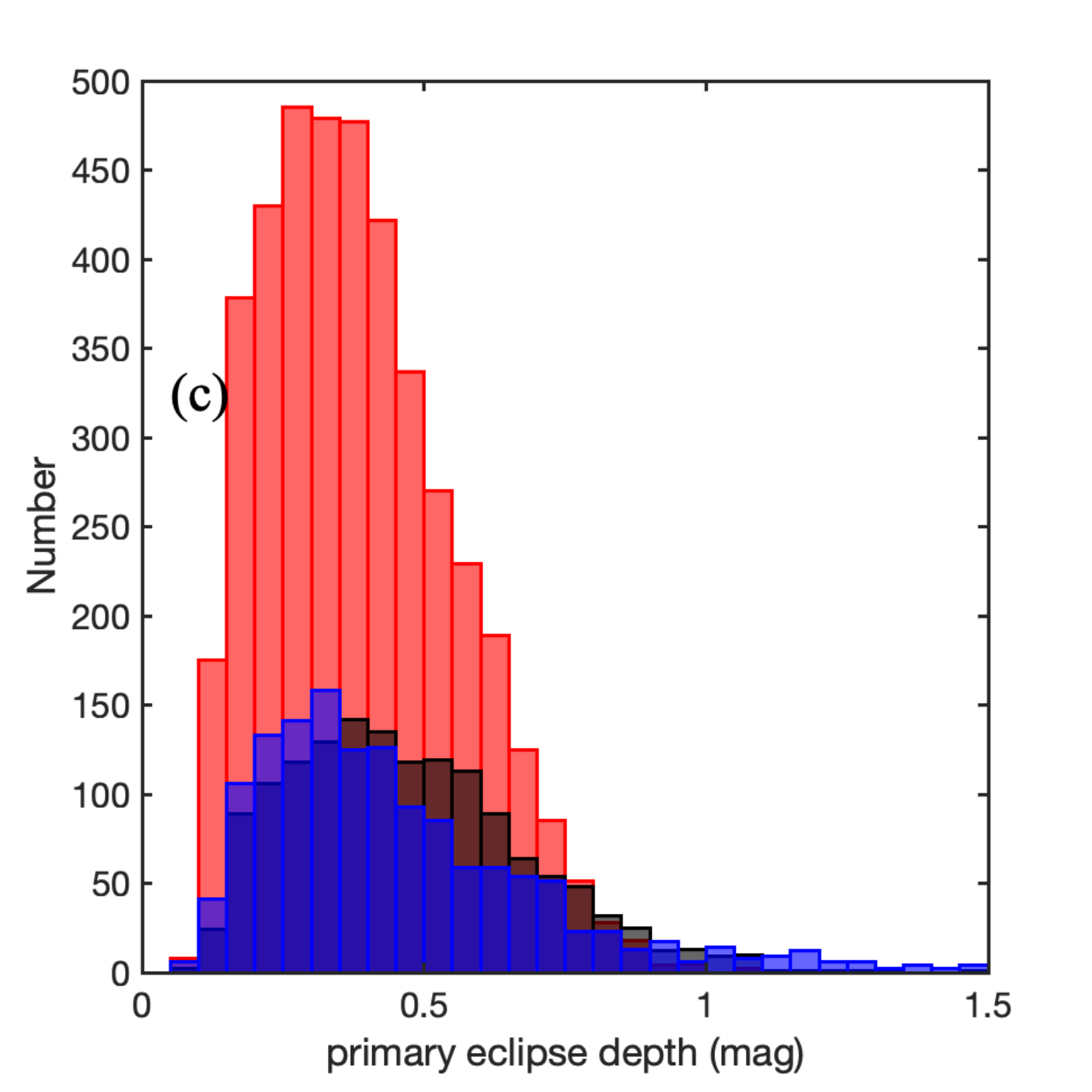}
\end{minipage}
\begin{minipage}[t]{0.32\textwidth}
\centering
\includegraphics[width=1\textwidth]{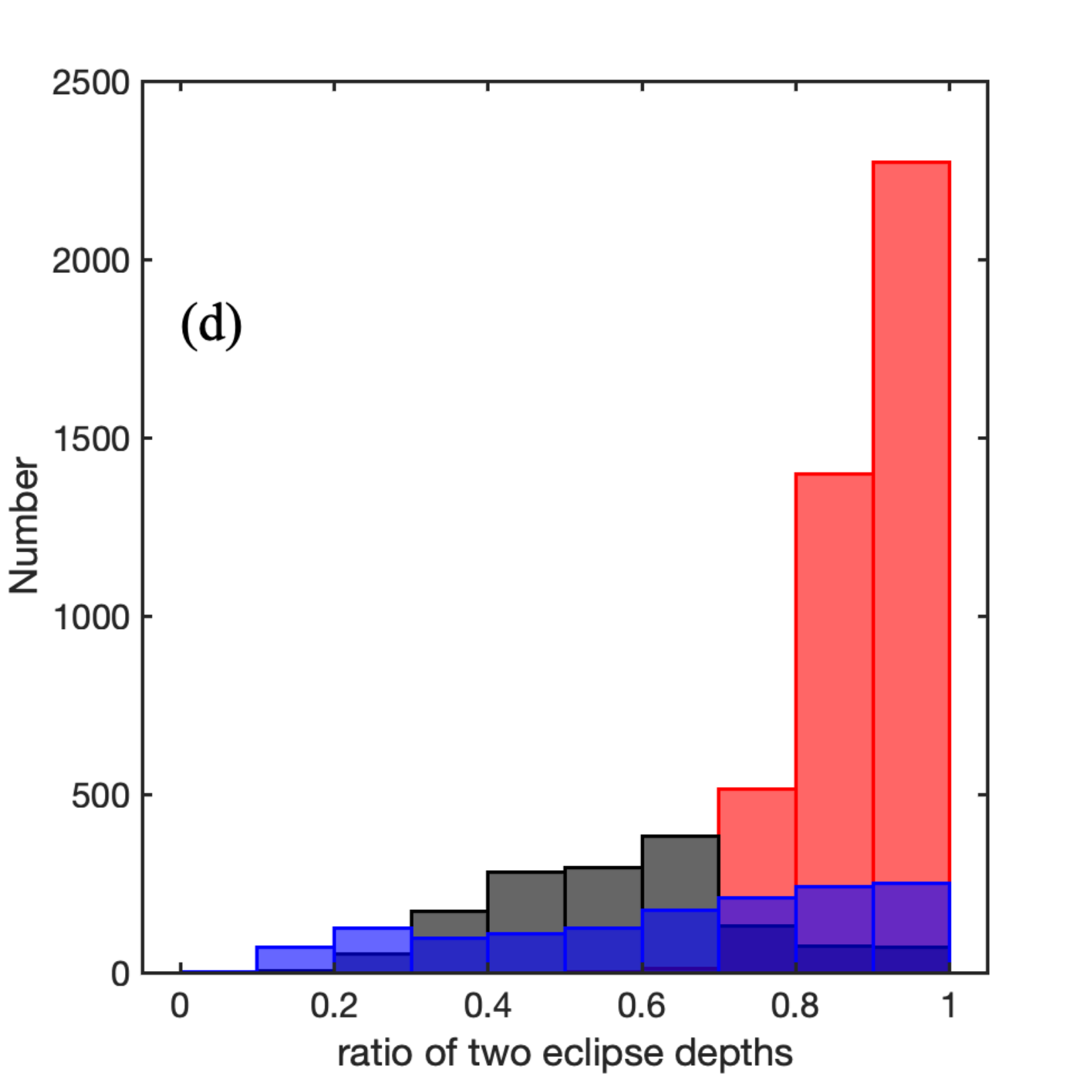}
\end{minipage}

\begin{minipage}[t]{0.32\textwidth}
\centering
\includegraphics[width=1\textwidth]{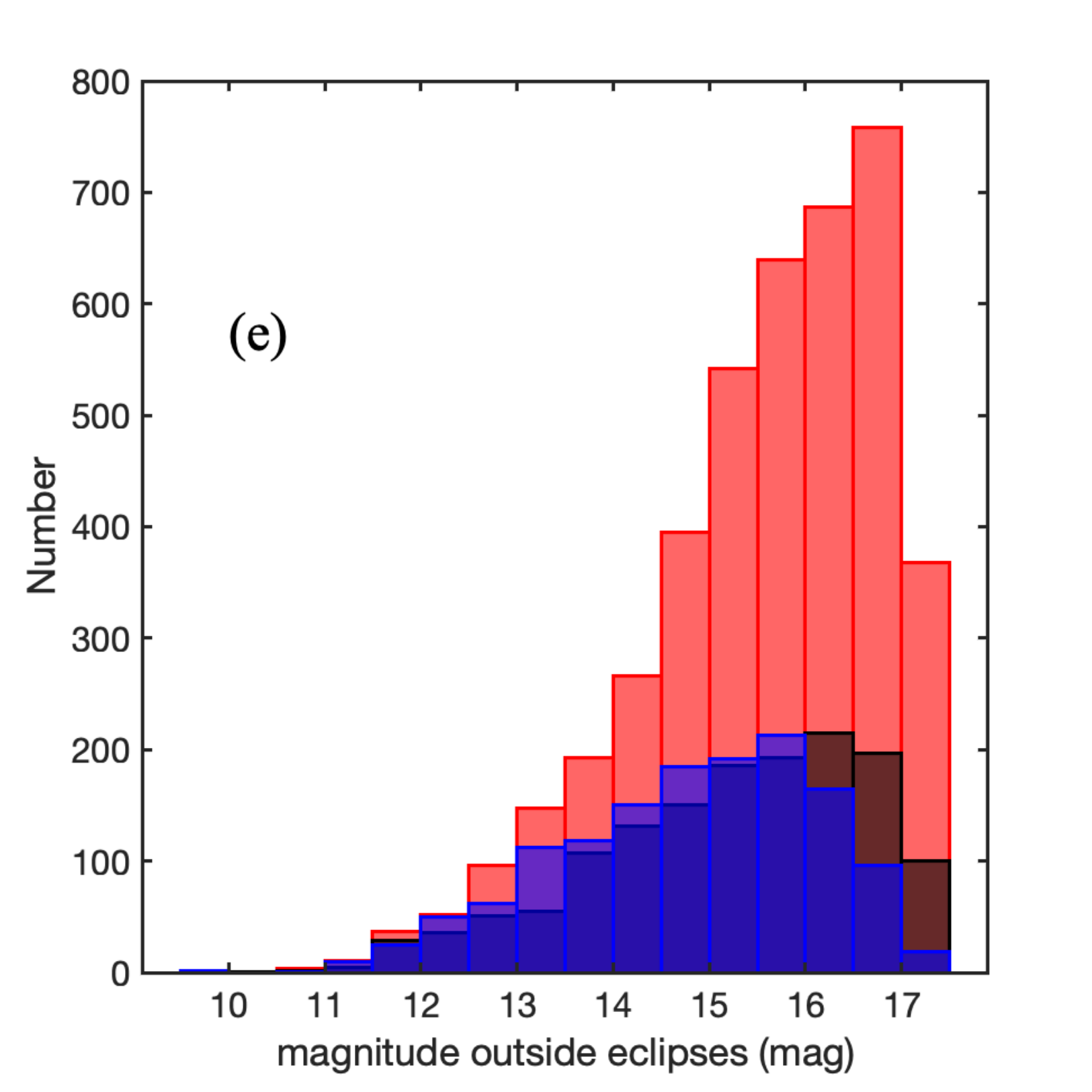}
\end{minipage}
\begin{minipage}[t]{0.32\textwidth}
\centering
\includegraphics[width=1\textwidth]{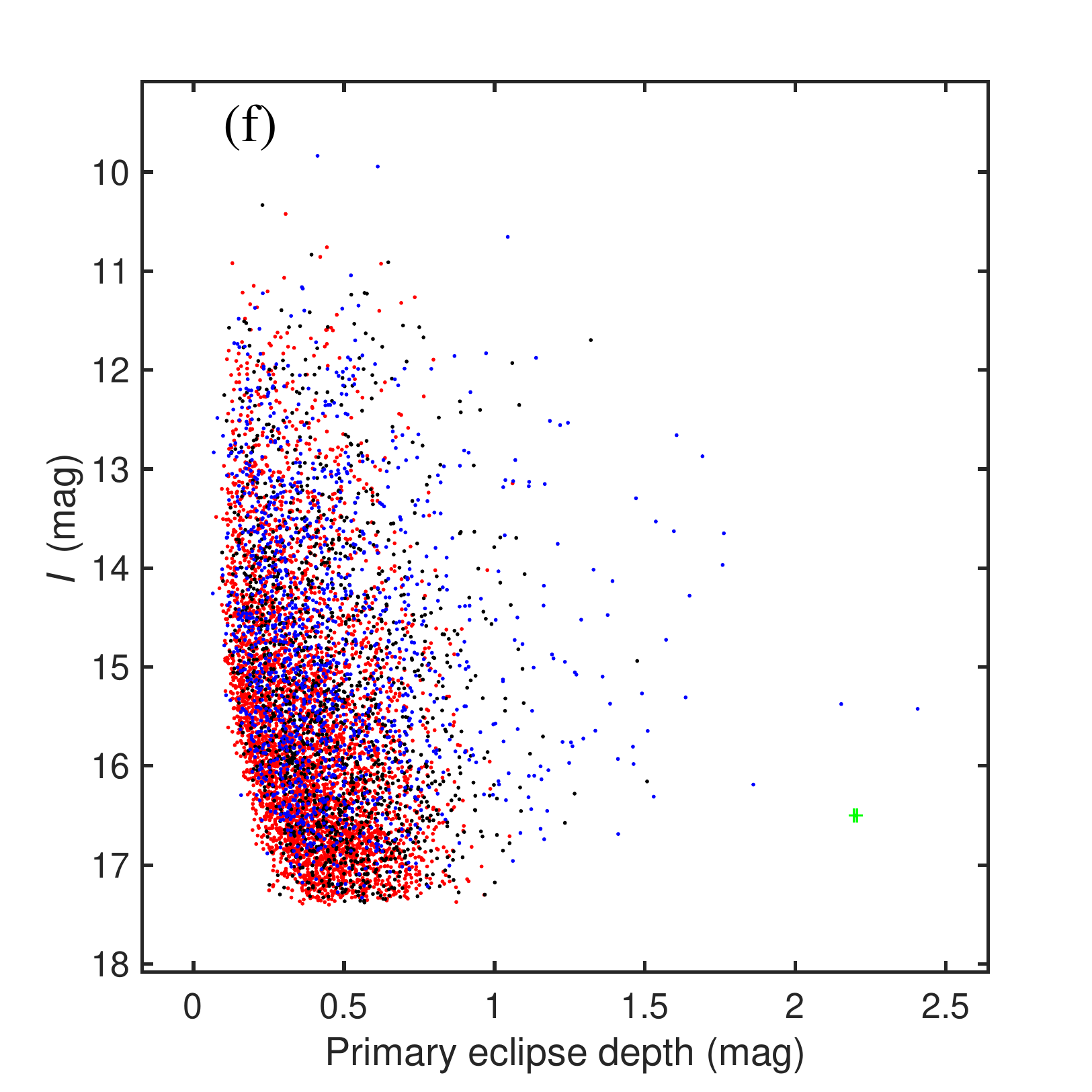}
\end{minipage}
\begin{minipage}[t]{0.32\textwidth}
\centering
\includegraphics[width=1\textwidth]{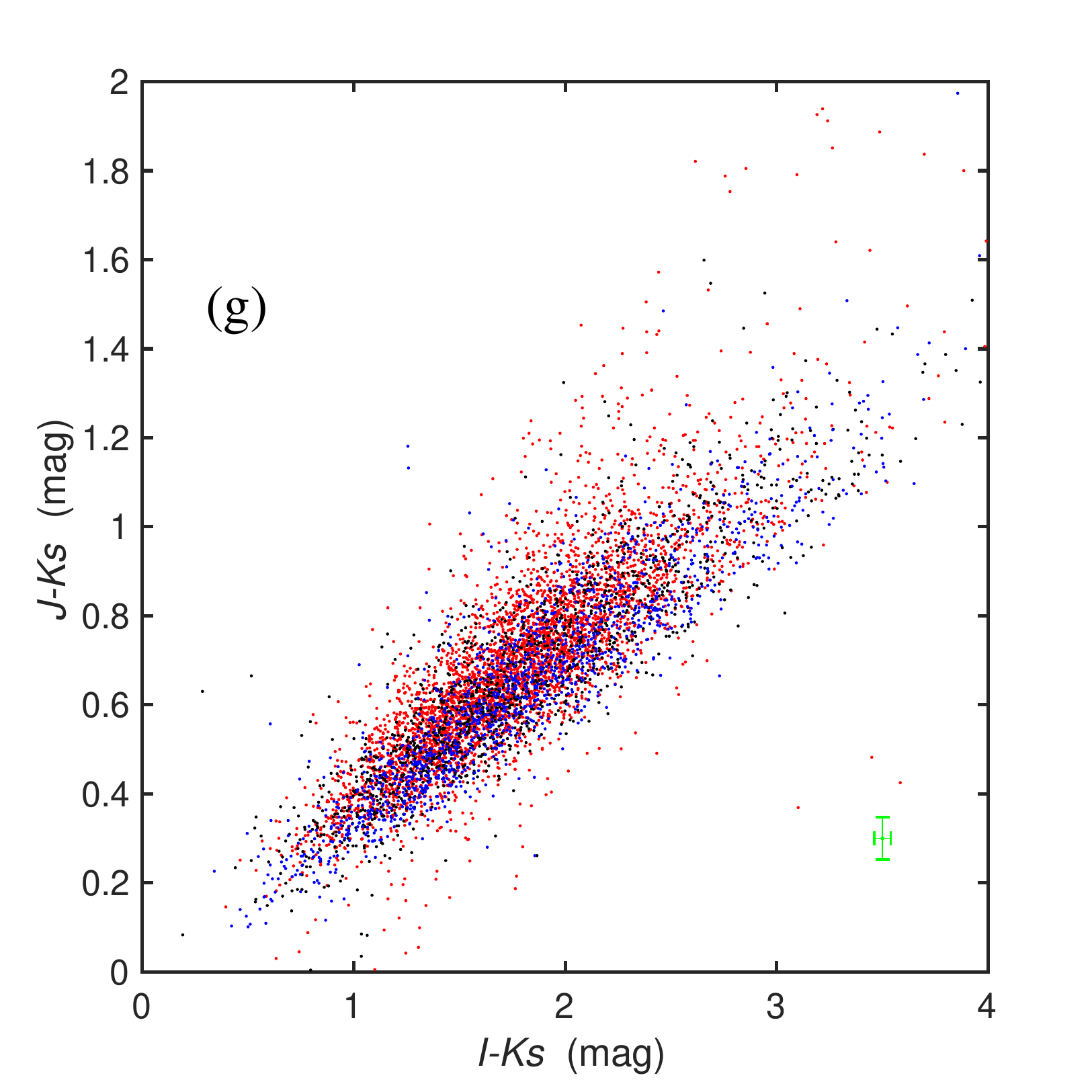}
\end{minipage}

\caption{EBS distributions. Red, black, and blue points and bars
  represent the 4197 EW-, 1458 EB-, and 1400 EA-type EBSs,
  respectively. The green points with error bars in panels (f) and (g)
  indicate the representative error bars pertaining to all samples in
  the respective panels. (top) Positions of all EBS in Galactic
  coordinates (in units of degrees). (middle left) Periods
  (logarithmic units). (middle center) Primary eclipse depths ($I$
  mag). (middle right) Ratios of secondary to primary
  eclipses. (bottom left) Magnitudes outside eclipses (total EBS
  magnitudes in the $I$ band). (bottom center) Scatter of the total
  $I$-band magnitude over primary eclipse depth for each EBS. (bottom
  right) Color--color diagram ($I-K_{\rm s}$ vs $J-K_{\rm s}$; with
  $J$ and $K_{\rm s}$ magnitudes from 2MASS).
\label{fig:distribution}}
\end{figure}

The parameter distributions for all EBS are shown in Figure
\ref{fig:distribution}. Red points or bars represent EW types,
while black and blue symbols correspond to EB and EA types,
respectively.

The EBS distribution as a function of Galactic longitude is shown in
Figure \ref{fig:distribution}{a}. It is inhomogeneous in number
density. This inhomogeneity is caused by the presence of groups of
stars associated with the Galaxy's spiral arms, including the Orion
Spur and the Perseus Arm. Since EBS variability is caused by the
systems' geometry, independently of any other criteria associated with
object position or selection bias, the EBS distribution is
representative of the general distribution of stars in the Galaxy
\citep{2006MNRAS.368.1319R}. Extinction also plays an important role
in the observed inhomogeneity, since dust obscuration and scattering
have a significant effect on stars in the Galactic plane. Only nearby
stars can be seen projected toward the most obscured regions. We will
return to the inhomogeneous distribution of EBS in Section
\ref{sec:structure}.

There are obvious period differences among the three types of EBS in
Figure \ref{fig:distribution}{b}. EA types have the longest periods,
ranging from 0.3 to 10 days, with a peak at several days. EW types
have the shortest periods, i.e., shorter than one day; most EW types
have periods around 0.3--0.5 days. Both the period range and the peak
period of EB types are found in the middle of the three types. Their
period distribution also depends on their environment. EBS in the
Galactic plane are thought to be less evolved and have longer periods
than those outside the thin disk, since they are usually younger. This
is shown in the left panel of Figure \ref{fig:10}; see Section
\ref{sec:purity} for a discussion.

For EA types, extremely wide period ranges have been reported in the
literature \citep[e.g.][]{2011AcA....61..103G}, reaching up to
thousands of days. Our detection approach prevents us from detecting
such long-period variables. We are constrained to a longest period of
50.22 days. For EA types with longer periods, it will be harder to
fully cover the eclipses to identify their types. However, instead of
continuous monitoring, only a few epochs are taken during a given
night for each object in our survey, which makes it harder to
completely cover their full eclipses. Some EA-type candidates with
extremely long periods may only have a few detections during their
eclipses. The numbers of detections may be insufficient to
characterize such an eclipse, and so such samples would fail to pass
our selection criteria. Therefore, our catalog lacks long-period
EA-type EBS.

No clear differences in the distribution of the primary depths can be
seen, but the distribution of the eclipse-depth ratios varies
significantly among the different EBS types (see Figure
\ref{fig:distribution}{c-d}). A clear plateau can be seen in the
histogram of the eclipse-depth ratios of EA-type EBS between 0.1 and
1. Secondary eclipses may not be visible if the stars are considerably
different in size. It is impossible to distinguish EA types with an
invisible secondary eclipse from EBS featuring a double period
characterized by similar eclipses solely based on light curves. This
may be the reason for the lack of a plateau at the shorter end of the
eclipse-depth distribution. Nevertheless, the observed distribution
underscores that the EA-type eclipse-depth ratios are randomly
distributed, while the ratio of EW types is close to unity, as
expected.

The histogram of the magnitudes outside eclipses, the scatter in
magnitude during the eclipses, and a color--color diagram exhibiting a
roughly linear relation for all types of EBS are shown in Figure
\ref{fig:distribution}{e-g}. The lack of faint EBS with shallow
eclipses is caused by selection effects. Faint systems with shallow
eclipses, whose magnitude errors during every single epoch are
comparable to the eclipse depth, would not pass our selection
procedure.

\subsection{Possible Contamination by Other Variables}\label{sec:purity}

We selected our EBS sample using the same method as that proposed
by \citet{2020ApJS..249...18C}. Contamination of our sample by
non-binary objects is less than 1\%, based on careful visual
examination. Since different types of variables exhibit different
kinds of light curves, it is easy to distinguish EA- and EB-type EBS
from other types of variables, even without access to any color
information. RRc Lyrae and EW-type EBS have similar periods and
amplitudes. An RRc Lyrae light curve with a nearly symmetric
luminosity decrease may be misclassified as an EW-type EBS
characterized by two similar-depth eclipses at double the RR Lyrae's
period. As such, RRc Lyrae are the main possible contaminant. Here
we evaluate the likely level of this type of contamination.

\begin{figure}[htbp]
\centering
\begin{minipage}[t]{0.64\textwidth}
\centering
\includegraphics[width=1\textwidth]{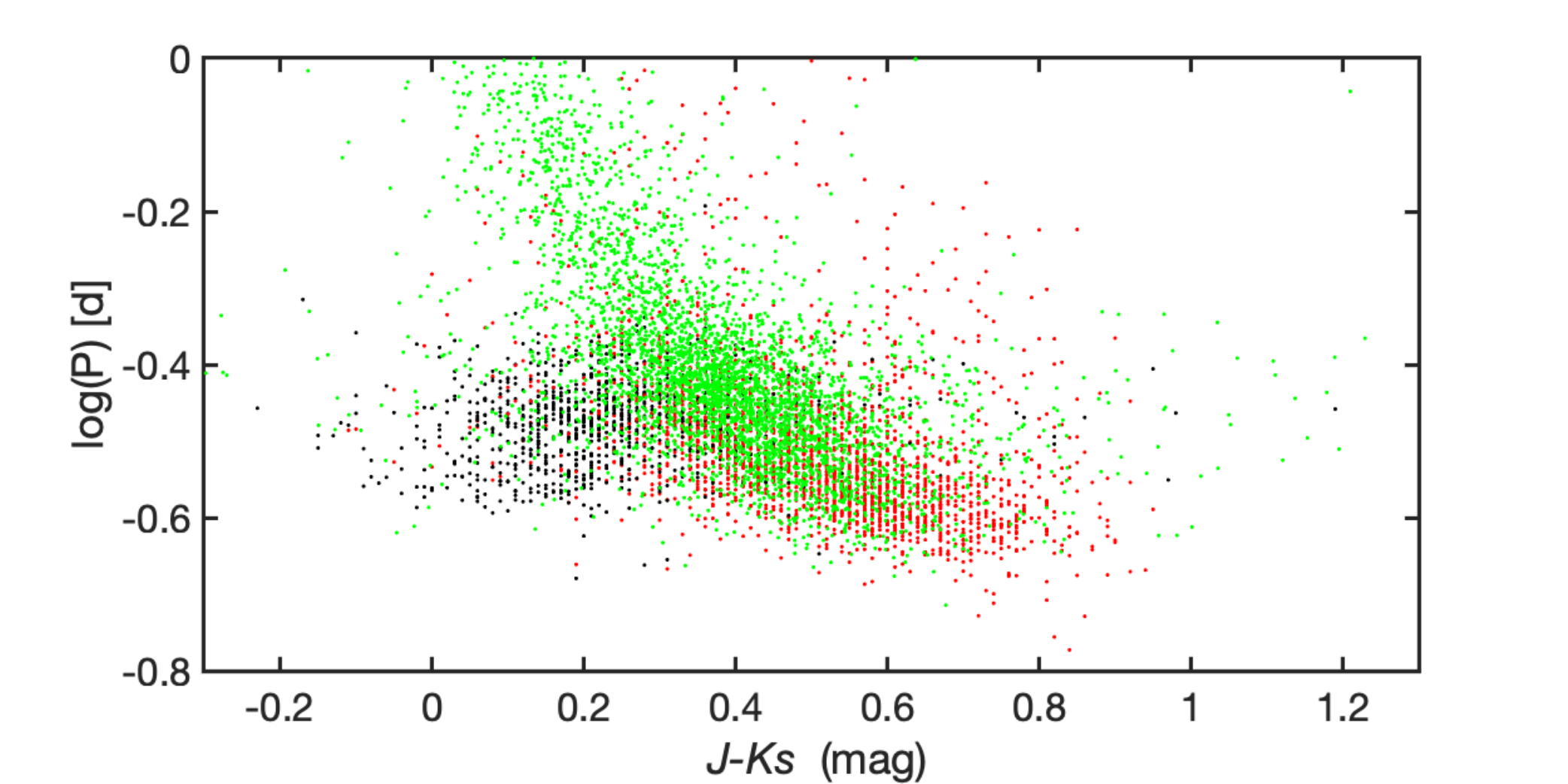}
\end{minipage}
\begin{minipage}[t]{0.32\textwidth}
\centering
\includegraphics[width=1\textwidth]{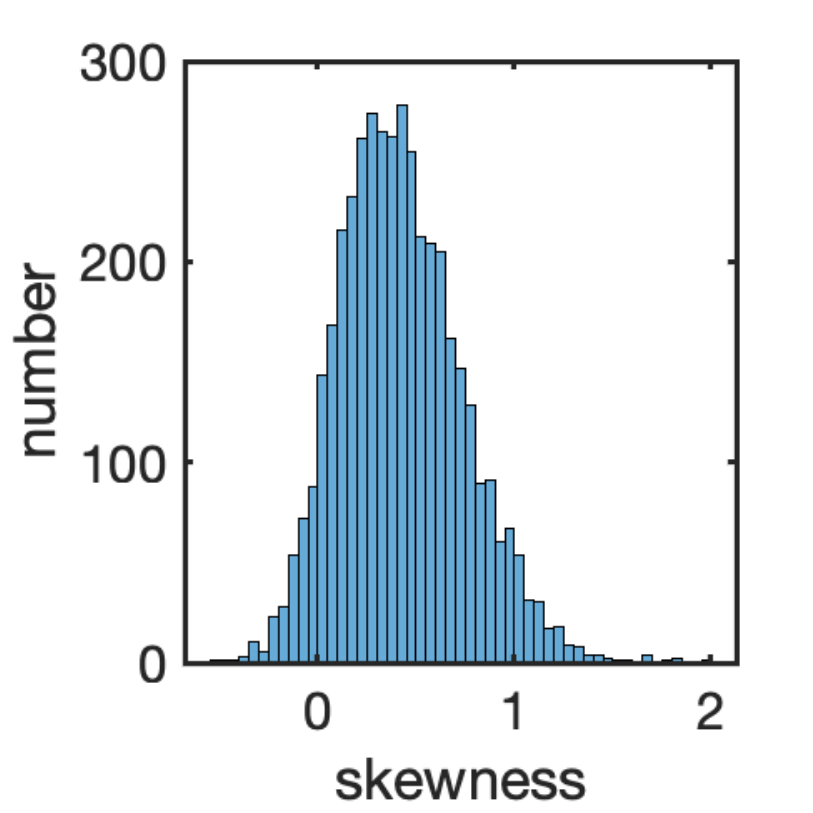}
\end{minipage}
\caption{Assessment of the possible contamination caused by RRc
  Lyrae. (left) EW-type color--period distribution. Green: Our EW-type
  sample objects; black and red: RRc Lyrae and EW-type systems,
  respectively, from \citet{2013AJ....146..101P}. (right) Skewness
  distribution of our sample. \label{fig:10}}
\end{figure}

This degeneracy can be lifted by considering the light-curve skewness
and color. EW-type objects are associated with a significantly larger
skewness than RRc Lyrae. The latter exhibit more sinusoidal-like light
curves, while EW-type variables remain brighter for longer during
their out-of-eclipse phase. EW-type objects are also significantly
redder. These properties will be helpful in aiding our efforts to
prevent contamination of our EW-type sample by RRc Lyrae \citep[see,
  e.g.][]{2013AJ....146..101P}.

Figure \ref{fig:10} presents the color--period and skewness
distributions of the EBS discussed in this paper. In the left panel,
we show the de-reddened 2MASS $J-K_{\rm s}$ colors as green points
(the extinction values were taken from Section \ref{sec:extinction}),
as well as the LINEAR RRc Lyrae and EW types (black and red points,
respectively). Since all LINEAR objects are located at high
Galactic latitude ($b > 29$\arcdeg), their extinction values are
negligible in $J-K_{\rm s}$. Among our LINEAR objects, RRc Lyrae are
generally bluer than EW-type EBS. Our EW-type sample objects are
scattered from the top to the right section of the color--period
plane, where we see little contamination by RRc Lyrae; they match the
LINEAR EW types well. The skewness distribution is shown in the right
panel. The adopted boundaries of the skewness, based on the LINEAR
sample, are $-0.1$ to 1.6 for EB/EW types and $-0.4$ to 0.35 for RRc
Lyrae. Our sample's distribution matches that of the LINEAR sample
quite well.

Although it is hard to exclude all RRc Lyrae solely based on our
current photometric data, contamination effects appear negligible
according to these color and skewness tests.

\section{Discussion of our EW-type sample objects}\label{sec:EWs}

\subsection{Extinction and Distance}\label{sec:extinction}

\citet{1968ApJ...151.1123L,1968ApJ...153..877L} proposed convective
common-envelope evolution as the key idea underlying EW-type theory.
The modern model \citep{2006AcA....56..199S,2013MNRAS.430.2029Y}
suggests that EW types form through both angular-momentum loss and
nuclear evolution. EW types are in their late stage of evolution, and
it has been shown observationally that both components have similar
temperatures. EW-type binaries usually form a common envelope and they
can thus be regarded as a single structure. In the color--magnitude
diagram, they appear as objects close to the main sequence (MS),
although one or both components may be more advanced in their
evolution
\citep{2006AcA....56..199S,2006AcA....56..347S,2009MNRAS.397..857S,2011AcA....61..139S}. Among
stars in the Galactic disk, EW-type objects are usually found in old
open clusters. Meanwhile, as one of the most numerous types of
variables in the Galactic field, they can potentially be used to
determine the Galactic disk's structure above and below the Galactic
plane and to trace any age gradient across the plane
\citep{2018ApJ...859..140C}. However, studies of the distribution of
EW-type systems in the Galactic plane are limited. With our sample, we
can now partially fill in the blanks.

Since the publication of \citet{1967MmRAS..70..111E}, a number of
attempts have been made to use EW-type period--luminosity--color
relations as potential distance indicators. Rucinski made several
attempts to derive PLRs from nearby EW types
\citep{1994PASP..106..462R}, EW types with {\sl Hipparcos} parallaxes
\citep{1997PASP..109.1340R}, ASAS catalogs
\citep{2006MNRAS.368.1319R}, and TGAS parallaxes
\citep{2017AJ....154..125M}. To unify the PLRs obtained from different
bands, optical-to-mid-infrared PLRs from \citet{2018ApJ...859..140C},
based on 183 nearby EW-type systems with TGAS parallaxes, were
adopted.

Note that our sample objects are located in the Galactic disk, where
the extinction is high and varies significantly. Although the KISOGP
survey can efficiently reduce the impact of extinction in the $I$ band
compared with that in $V$ band, extinction is expected to still have a
sizeable effect on the resulting photometry.

To derive accurate distances, we considered distance moduli ($\mu_0$)
and extinction values ($A_\lambda$ as a function of passband,
$\lambda$) as variable parameters, i.e.
\begin{equation}
m_\lambda-a_\lambda \times \log P-b_\lambda=(A_\lambda/A_V)\times A_V+\mu_0.
\end{equation}

For a given passband $\lambda$, we take the PLR coefficients
$a_\lambda$ and $b_\lambda$ from the maximum-magnitude coefficients of
the PLRs of \citet{2018ApJ...859..140C}; the extinction law,
$A_\lambda/A_V$, from \citet{2019ApJ...877..116W}; and the period,
$\log P$, from Table \ref{tab:1} in Section \ref{sec:deter}. For the
apparent magnitude, $m_\lambda$, we adopt the maximum $I$-band
magnitudes from Table \ref{tab:1}, combined with $J H K_{\rm s}$-band
data from 2MASS. However, 2MASS only provides single-epoch photometry,
as well as a time stamp for the observation. Since observed color
changes are very small during eclipses, the light-curve shape barely
changes among the different bands \citep{2016ApJ...832..138C}. This
makes it possible to convert the single-epoch magnitudes to the
corresponding maximum magnitudes obtained from the full $I$-band light
curves. The differences between the $I$-band magnitudes in the
observational phase and the maximum $I$-band magnitudes are listed in
Table \ref{tab:1}. These photometric properties also allow us to
convert the 2MASS single-epoch photometry to the maximum $J H K_{\rm
  s}$-band magnitudes.

\begin{figure}[ht!]
\centering
\includegraphics[width=8cm]{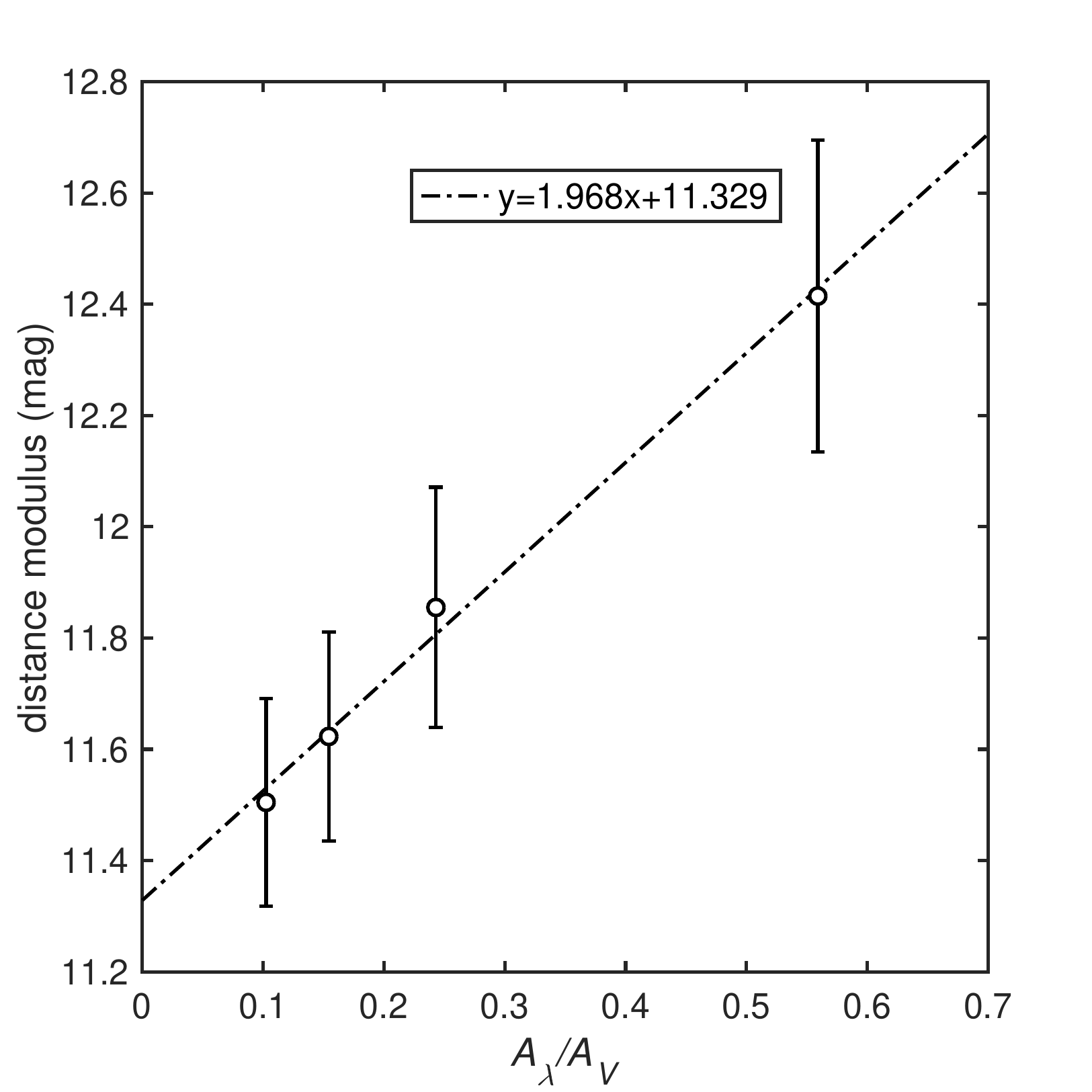}
\caption{Example of our determination of extinction and distance. The
  intercept corresponds to the distance modulus, $\mu_0 = 11.329$ mag,
  while the slope reflects the extinction in the $V$ band, $A_V =
  1.968$ mag. \label{fig:example}}
\end{figure}

Armed with known or newly determined parameters pertaining to the $I J
H K_{\rm s}$ bands, it is now possible to fit the $(\mu_0,A_V)$
combination for each system; for an example, see Figure
\ref{fig:example}. A similar method has been adopted by other
  authors \citep[e.g.][]{2017ApJ...842...42M}.

In this example, the slope corresponds to the $V$-band extinction,
$A_V$, and the intercept is the best-fitting distance modulus,
$\mu_0$. Since the PLR was fitted only to EW-type objects, our results
are based on 3996 objects, having first excluded systems without 2MASS
observations. The best-fitting distances and extinction values are
listed in Table \ref{tab:1}. For some sources affected by low
extinction, negative slopes implying negative extinction may result
for numerical reasons; in those cases we set the extinction to zero.

As an independent means to determine distances, EW-type PLRs have a
wide variety of uses, including as a probes to map the structure of
the Galactic plane and as benchmarks to crosscheck other means of
distance determination.

\subsection{Mapping the Galactic Thin Disk Structure with EW types}\label{sec:structure}

Equipped with distance estimates from our PLR application, it is now
feasible to derive the structure of the northern Galactic plane as
traced by EW-type EBS. To minimize selection effects, we excluded
KISOGP objects located at $70\arcdeg<l<80\arcdeg,1\arcdeg<b<3\arcdeg$
to retain a uniform distribution across $-1\arcdeg<b<1\arcdeg$ of the
thin disk.

\begin{figure}[htbp]
\centering
\begin{minipage}[t]{0.6\textwidth}
\centering
\includegraphics[width=1\textwidth]{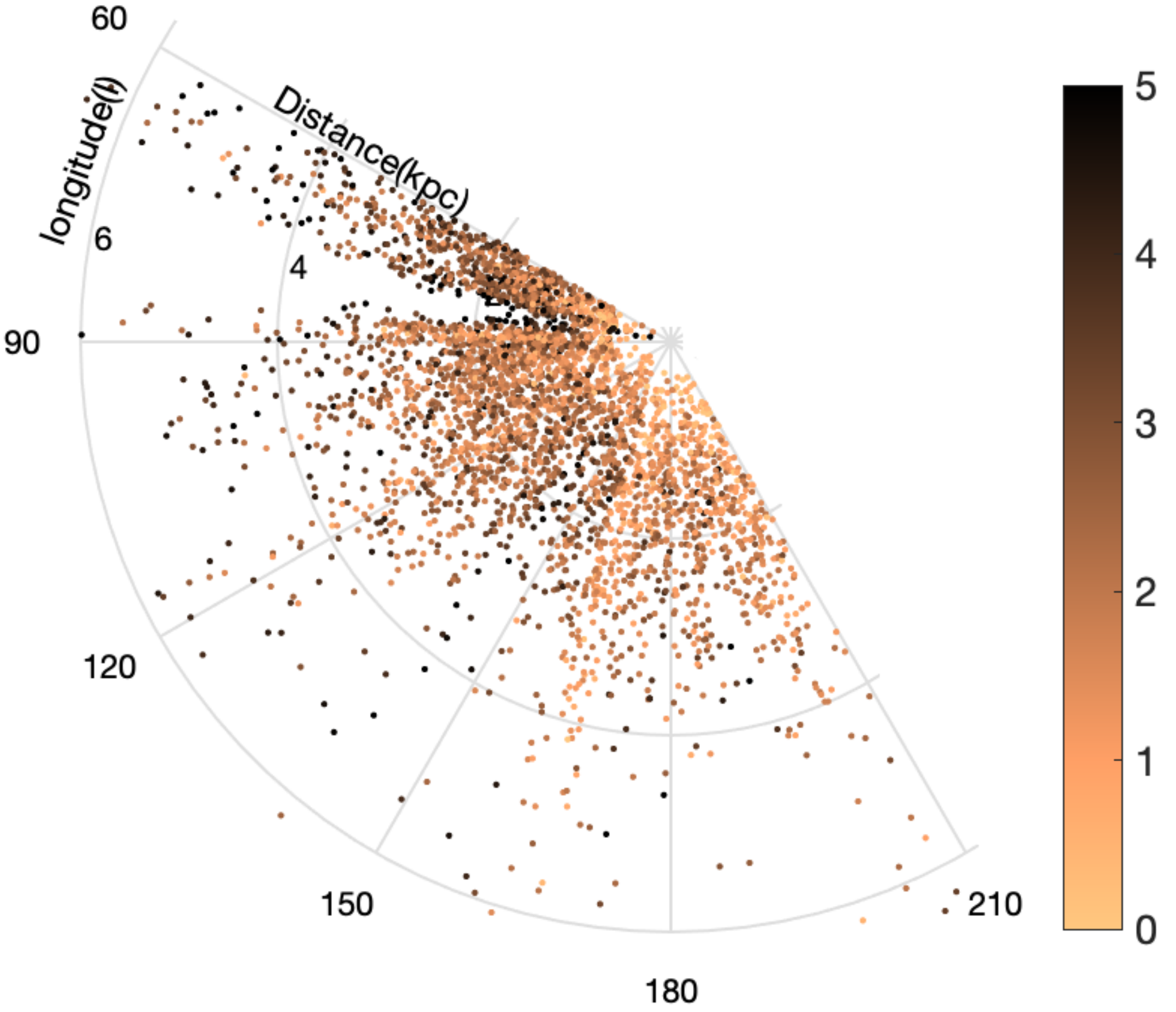}
\end{minipage}
\begin{minipage}[t]{0.7\textwidth}
\centering
\includegraphics[width=1\textwidth]{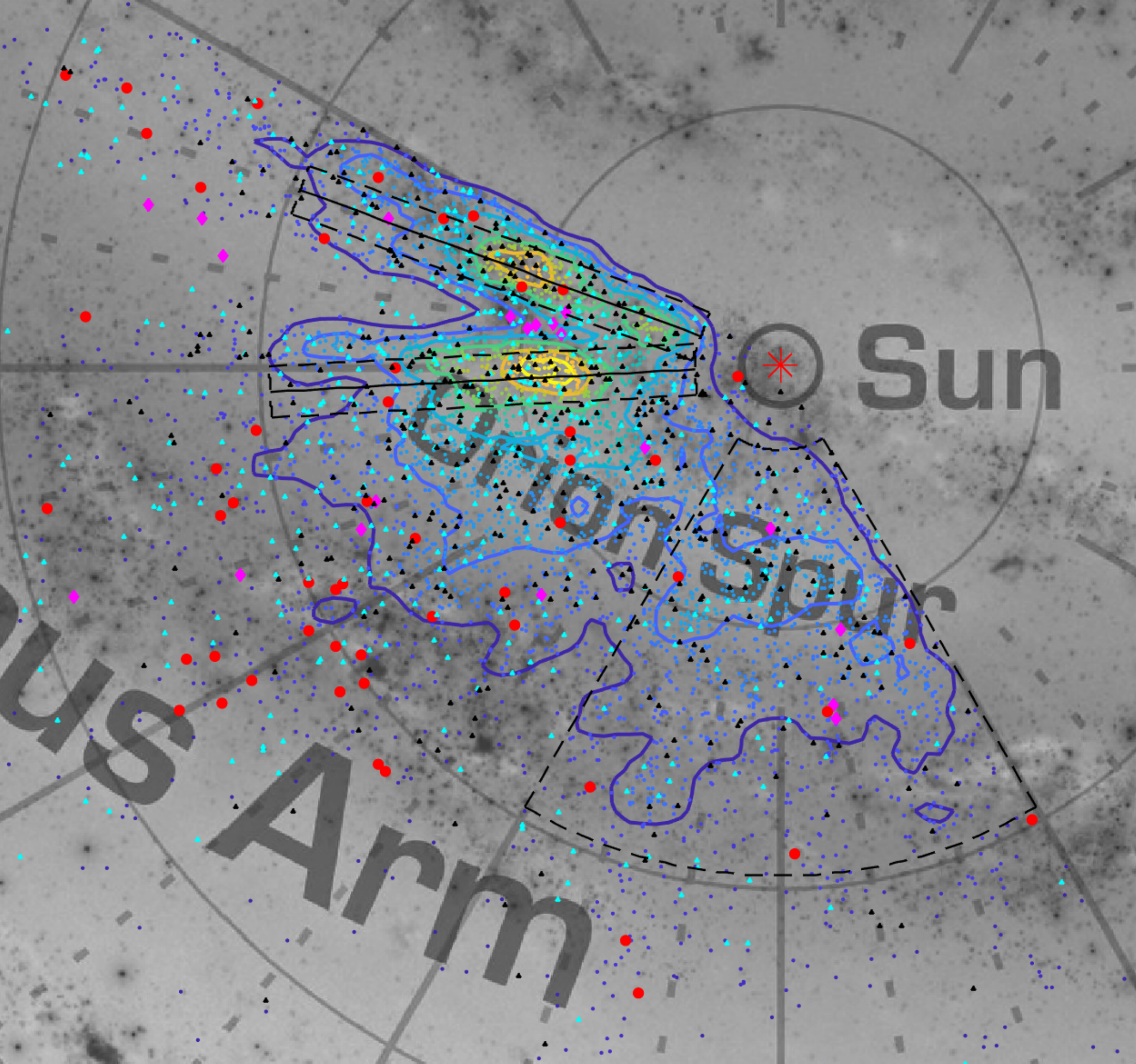}
\end{minipage}
\caption{Spatial distribution of EW-type EBS in the Galactic thin
  disk. (top) Space and extinction distributions looking down on the
  Milky Way, centered on the Sun. Different colors represent different
  extinction values as reflected by the color bar. (bottom) Spatial
  distribution of the EW-type EBS as seen from the Galactic North
  Pole. A density map of our EW types is shown as contours (from blue
  to yellow). The Sun is shown as a red star symbol. Some other
  tracers are also included for comparison. High-mass star-forming
  regions \citep{2014ApJ...783..130R} are shown as magenta diamonds,
  Galactic Cepheids \citep{2014A&A...566A..37G} are shown as red
  circles, while EA- and EB-type EBS from Table \ref{tab:1} with
  accurate {\sl Gaia} parallaxes ($0<\sigma_\pi/\pi<0.1$) are shown as
  cyan and black triangles, respectively. Members of all groups are
  dispersed across the northern Galactic thin disk in the same
  region. Background image credit: Modified from the original artist's
  conception; NASA/Joint Propulsion Laboratory--California Institute
  of Technology/R. Hurt (Spitzer Science Center).
\label{fig:strcture}}
\end{figure}

The spatial and extinction distributions of EW-type tracers derived
from application of the PLR are shown in the top panel of Figure
\ref{fig:strcture}. Different colors represent different extinction
values; all objects affected by $A_V\ge5$ mag have been assigned the
same color. Our sample objects near the Sun are, in general, affected
by low extinction, usually less than $A_V=2$ mag, while objects in
more distant regions are significantly more highly obscured. The
spatial distributions defined by different tracers, as well as that
traced by our EW types, are shown in the bottom panel of Figure
\ref{fig:strcture}. Members of high-mass star-forming regions
\citep[HMSFRs;][]{2014ApJ...783..130R}, Galactic Cepheids
\citep{2014A&A...566A..37G}, and the 478 EA- and 454 EB-type
objects from Table \ref{tab:1} with accurate {\sl Gaia} parallaxes
($0<\sigma_\pi/\pi<0.1$) are shown as magenta diamonds, red circles,
and cyan and black triangles, respectively. EW-type systems with
distances obtained from the PLR are also shown in the density map. The
density map was constructed using MathWorks' `scatplot'
tool\footnote{\url{https://www.mathworks.com/matlabcentral/fileexchange/8577-scatplot}},
while colored data points and contours were plotted using Voronoi
cells to determine the relevant densities. Members of all groups are
dispersed in the same region across the northern Galactic thin disk.

The density distributions of EW-type and other tracers show that the
Galactic thin disk appears inhomogeneous. This exercise can
potentially help us to reveal substructures such as bubbles and
filaments. Meanwhile, the components of EBS usually form from MS
stars, and their differences within each type of EBS are not affected
by age (or at most to a limited extent). Different EBS types are
generally associated with different ages, in the sense that EA types
are younger, EB types are older, and EW types are the oldest tracers.
Meanwhile, Cepheids are young stars and HMSFRs are very young. These
different tracers can thus also be used as age tracers.

The derived structure of the thin disk varies significantly as a
function of direction. In some directions, e.g. toward
$l\sim120\arcdeg$ and $l\sim165\arcdeg$, EW types can be seen out to
considerable distances given the slowly increasing extinction trends
there. It appears that these directions offer low-extinction windows,
reaching and even crossing the Perseus Arm. These areas are known as
diffuse regions \citep[][their Figure 1]{Wang_2017}. In other
directions, the extinction increases quickly and the largest visible
distance from the Sun is small. For example, around $l\sim80\arcdeg$ a
clear lack of objects is caused by high extinction, $A_V > 5$ mag
within 1 kpc. A young, dense cloud with high star-forming activity and
high extinction is located in that direction \citep[i.e., the
Cygnus X region;][]{2012A&A...539A..79R}. Its location among the
accumulation of EBS may have been caused by second-generation
star-forming activity. The double-peaked features along the sightlines
to $l\sim70\arcdeg$ and $93\arcdeg$ are caused by significant
extinction toward $l\sim80\arcdeg$, which prevents detections of
objects. The Orion Spur happens to be located in this dense region
(from a distance of 3 kpc toward $l\sim70\arcdeg$ to 2 kpc toward
$l\sim93\arcdeg$; for reference, the distance of the equidistant
circle in the background image is 5000 light-years or about 1.5
kpc). Similarly rapidly increasing extinction values can also be seen
toward $l\sim145\arcdeg$. Unlike the line of sight toward
$l\sim80\arcdeg$, star-forming activity is not found here.

\begin{figure}[htbp]
\centering
\begin{minipage}[t]{0.24\textwidth}
\centering
\includegraphics[width=1\textwidth]{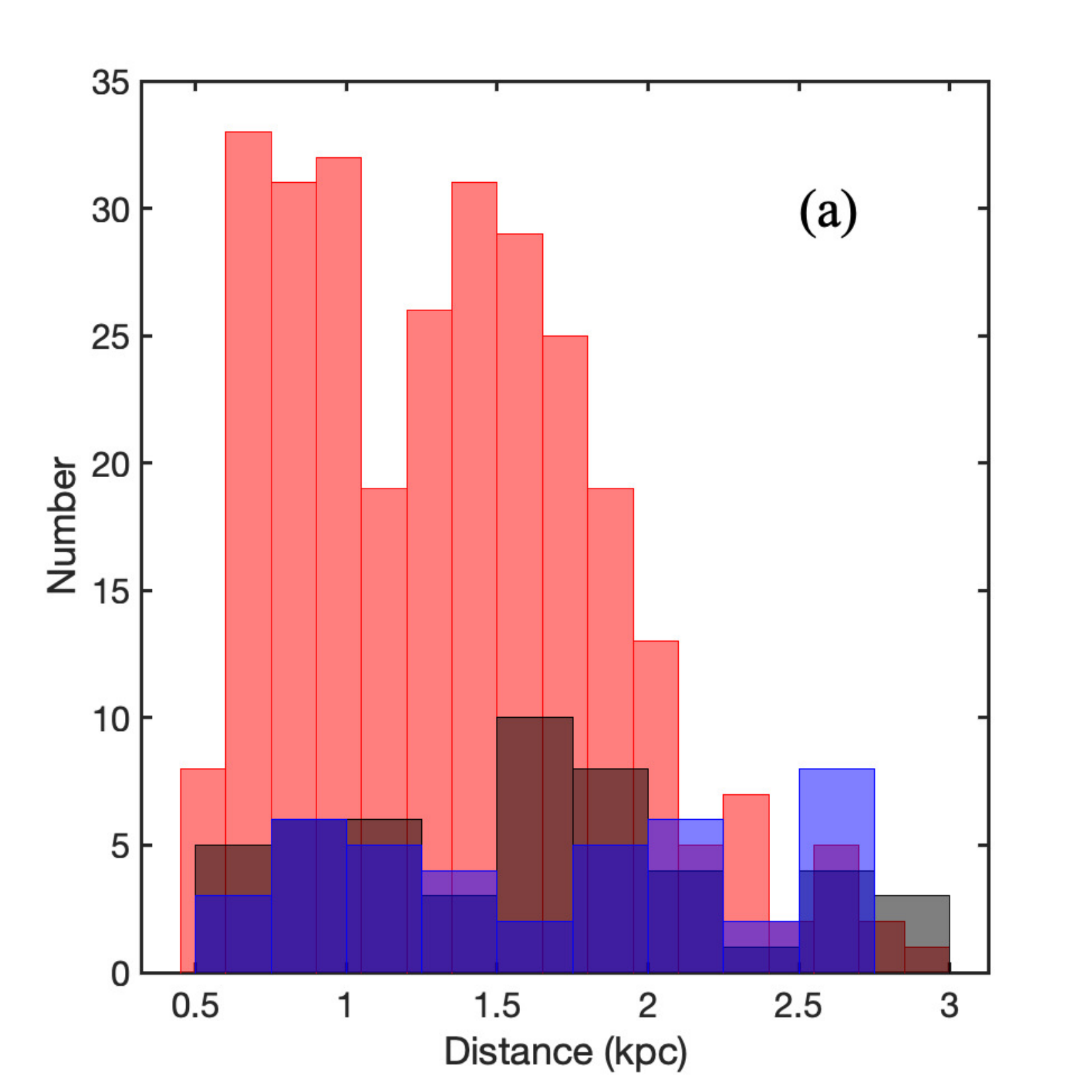}
\end{minipage}
\begin{minipage}[t]{0.24\textwidth}
\centering
\includegraphics[width=1\textwidth]{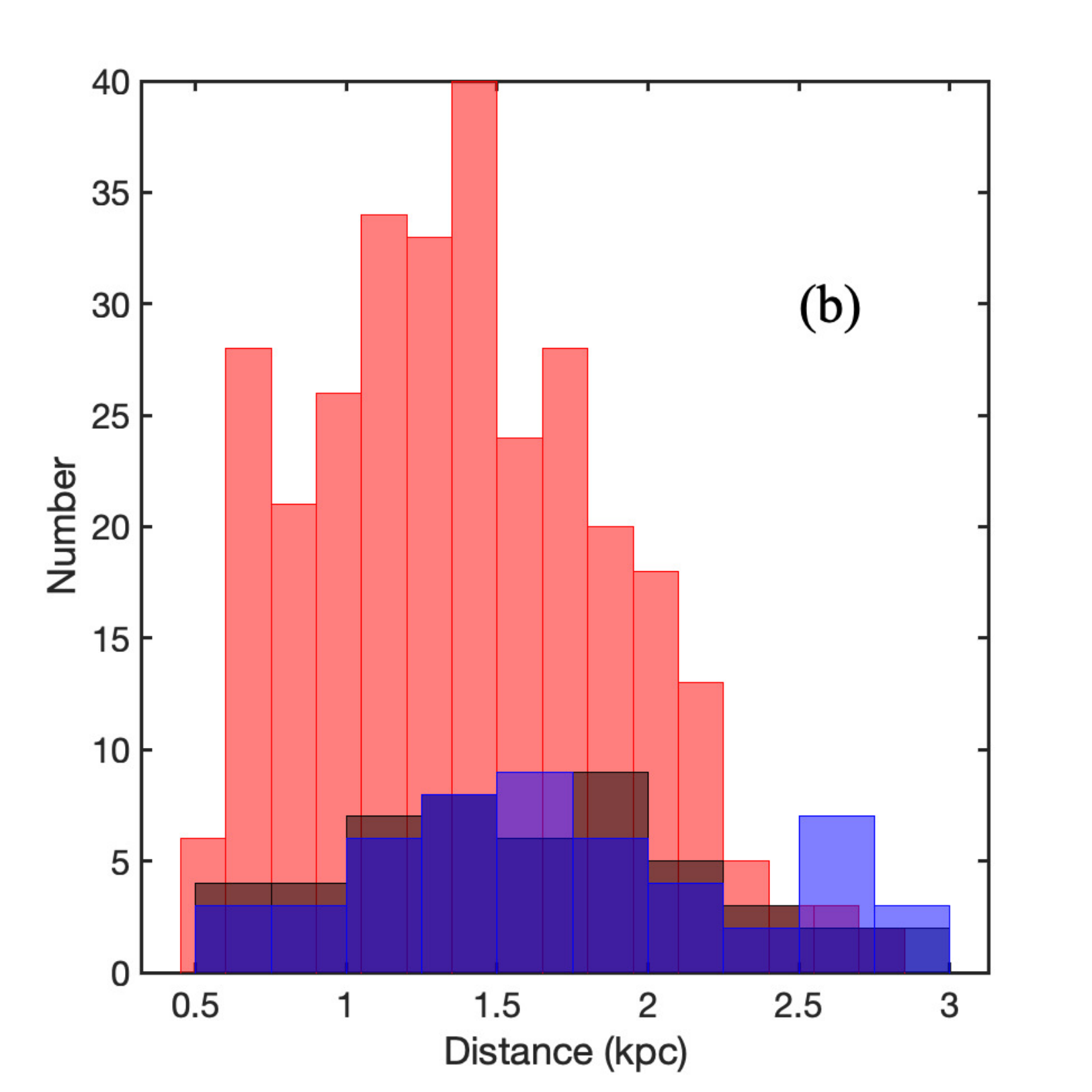}
\end{minipage}
\begin{minipage}[t]{0.24\textwidth}
\centering
\includegraphics[width=1\textwidth]{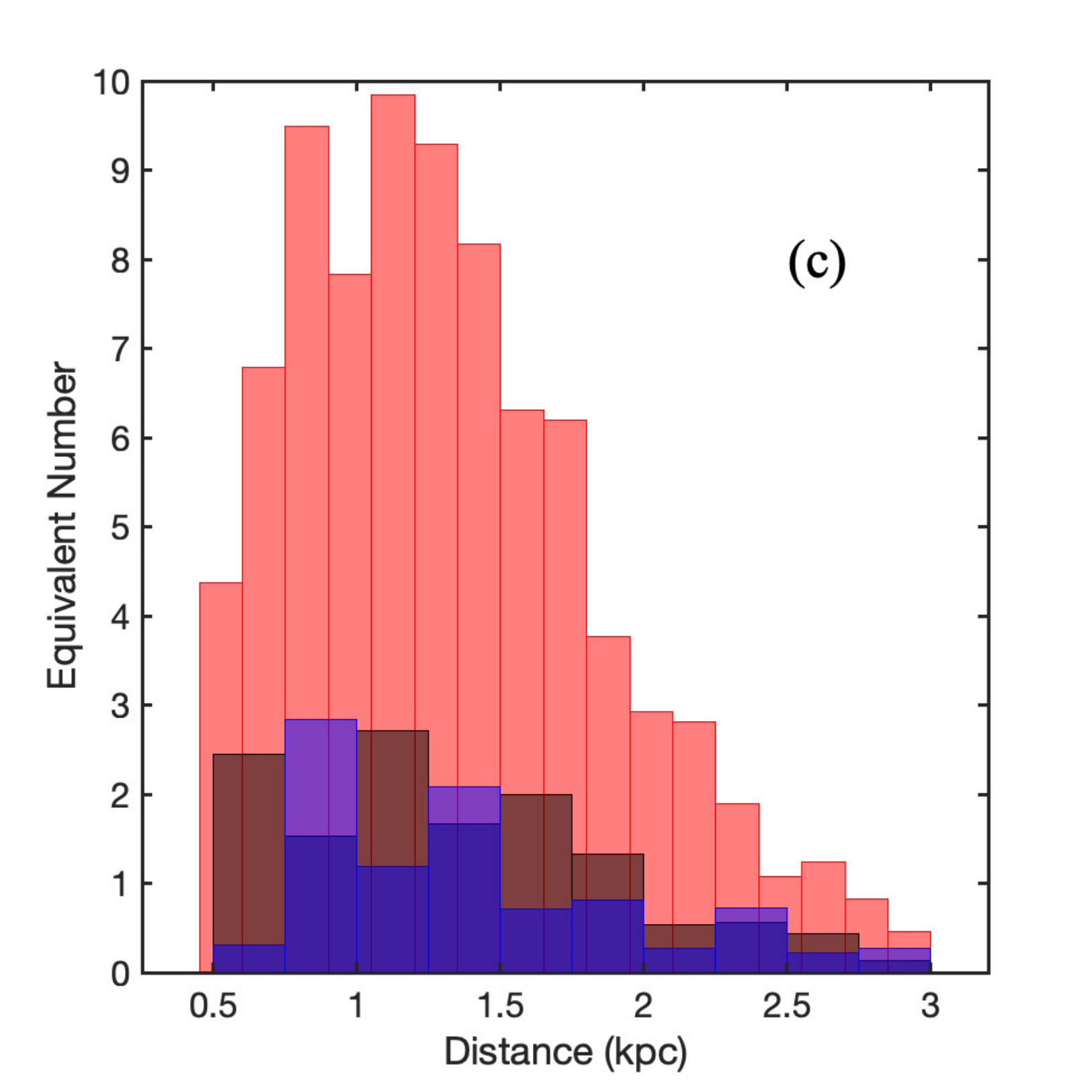}
\end{minipage}
\caption{ Detailed distributions of EBS in different distances
along three sightlines. Red, black, and blue bars represent EW-,
EB-, and EA-type EBSs, respectively. (a) $l=70$\arcdeg. (b)
$l=93$\arcdeg. (c) $l=150$--210\arcdeg. The equivalent numbers in
this latter panel were calculated from a combination of the
relevant number density and the area covered, which is the same as
for panels (a) and (b) at the same distance ($N_{\rm
  equi}=S_{a,b}\times \frac{N}{S_{c}}$). 
\label{fig:detailed}}
\end{figure}

In general, the stellar distribution appears inhomogeneous as a
function of direction along the Galactic plane. For example, for
distances between 0.5 kpc and 3 kpc, compare the sector between
$l\sim70\arcdeg$ and $l\sim93\arcdeg$ with a second sector between
$l=150$\arcdeg and $l=210$\arcdeg: see the bottom panel of Figure
\ref{fig:strcture}. The regions contained within the dashed
rectangles have widths of 300 pc. The corresponding EBS
distributions are shown in Figure \ref{fig:detailed}. Figure
\ref{fig:distribution}(e) suggests that our sample's completeness is
higher for systems brighter than $I = 16$ mag, an upper limit we
therefore adopt for our subsequent analysis.

Figure \ref{fig:detailed}a shows that the systems located along
this specific sightline ($l\sim70\arcdeg$) are distributed
inhomogeneously and tend to cluster on scales of several hundred
parsecs, irrespective of EBS type. The distribution of EW types
exhibits two ridges separated by a valley on a scale of approximately
1 kpc. The EB- and EA-type distributions also show two ridges, within
about 2.5 kpc, although the details differ. Within our target
sectors, the spatial clustering scale, i.e., the width of the nearby
ridges, increases from EW through EB to EA types. Along the same
direction, an age gradient can also be discerned. Moreover, the
extinction behavior also suggests that these clustering scales
reflect reality.

However, the sightline toward $l\sim93\arcdeg$ shows a different
pattern, as shown in Figure \ref{fig:detailed}(b). All three types
of EBS exhibit clustered distributions between 1 and 2 kpc, but only
EA types show a rise beyond 2.5 kpc. Compared with Figure
\ref{fig:detailed}(c), which also offers some evidence of clustering
behavior, the peak of the distribution along this sightline is
located at greater distances. We suspect that the peaked
distributions of our EBS may be driven by the structure of the thin
disk, as traced by EBS, rather than by sampling incompleteness

As such, it is clear that the EBS density distribution in the thin
disk is not uniform. Future work should address the detailed
structure of the Galactic thin disk based on a larger EBS sample.

\subsection{Absolute parameters}\label{sec:parameter}

To arrive at homogeneous estimates of the physical parameters of the
EW-type EBS in the Galactic plane, we created a model for each system
using the 2015 version of the Wilson--Devinney (W--D) code
\citep[][]{1971ApJ...166..605W, 1979ApJ...234.1054W,
  1990ApJ...356..613W,2020ApJS..247...50S}. The input parameters of
the models were based on the results listed in Table \ref{tab:1}. We
excluded systems with fewer than 150 photometric epochs given
the limited accuracy of the resulting parameters.

\begin{figure}[htbp]
\centering
\begin{minipage}[t]{0.48\textwidth}
\centering
\includegraphics[width=1\textwidth]{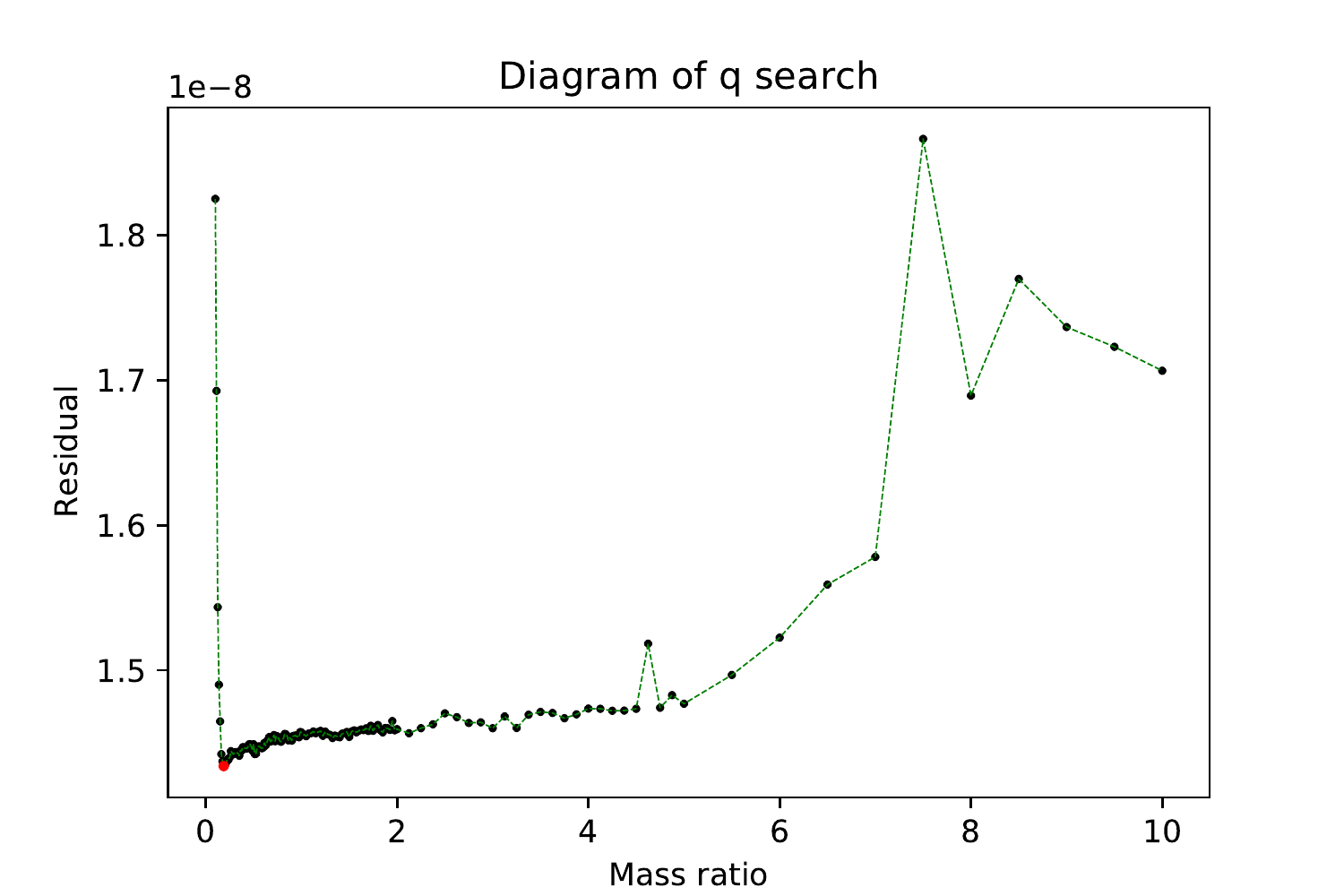}
\end{minipage}
\begin{minipage}[t]{0.48\textwidth}
\centering
\includegraphics[width=1\textwidth]{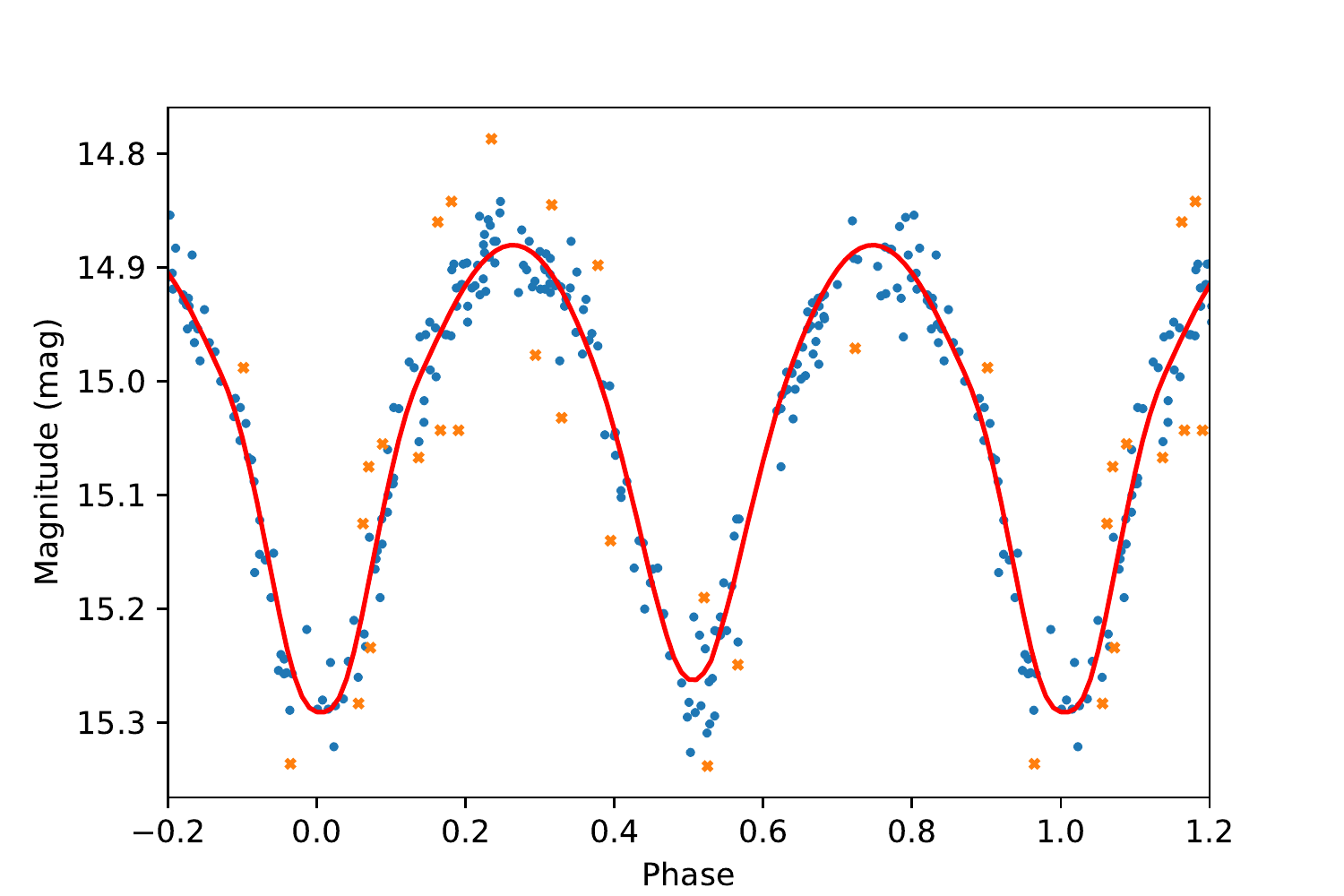}
\end{minipage}
\caption{Example of our parameter determination procedure. (left) Mass
  ratio search. The mass ratio associated with minimum residuals was
  adopted; here $q=0.1875$ (shown as a red dot). (right) Best-fitting
  light curve overplotted on the original observations ($I$-band
  magnitudes). The outliers are marked as orange
  crosses. \label{fig:q_search}}
\end{figure}

For all systems, `Mode 3' (contact mode, usually applied to systems in
geometric contact without constraints on the thermal contact
configuration) was used to analyze the light curves. The input light
curves were based on the parameters listed in Table \ref{tab:1}. The
effective temperatures of the primary components were taken from
\citet{2013ApJS..208....9P}, based on their 2MASS $(J-K_{\rm s})$
colors and the extinction calculated previously. The distances derived
from the PLR were used to render absolute system parameters. The
bolometric corrections are from \citet{2019A&A...632A.105C}.

EW types can be divided into two groups, split at a photospheric
temperature of 6200 K \citep{2017MNRAS.465.4678M}. The hotter objects
correspond to stars dominated by radiative energy transport while
cooler objects are stars with convective envelopes. We set the
gravity-darkening coefficients and the bolometric albedos to,
respectively, $g=0.32$ and $A=0.5$, and $g=1.0$ and $A=1.0$, for stars
with temperatures less and greater than 6200 K, respectively. A
bolometric logarithmic limb-darkening law was applied. No spots, third
bodies, or time derivatives of the orbital period were considered in
our light-curve fitting.

We used an extensive $q$-search method to find the best mass ratio,
$q=M_2/M_1$. Mass ratios $q$ from 0.1 to 10 were tried, in steps of
0.0125 from 0.1 to 1, 0.025 from 1 to 2, 0.125 from 2 to 5, and 0.5
from 5 to 10; see the left panel of Figure \ref{fig:q_search}. The
mass ratios are shown on the horizontal axis, while the residuals are
shown on the vertical axis. In this example, the minimum residual
occurs for a mass ratio $q=0.1875$. The best-fitting light curve,
compared with the original observations, is shown in the right panel
of Figure \ref{fig:q_search}. Outliers rejected during our light-curve
solution procedures are marked as orange crosses. The phase shift of
the primary eclipse, the orbital inclination ($i$), the temperature of
the secondary component ($T_2$), the potentials of both components
($\Omega_1$, $\Omega_2$; $\Omega_1=\Omega_2$ in the geometric contact
configuration), and the bandpass luminosities of the primary component
($L_1$) were treated as adjustable parameters. To arrive at a more
reliable sample, we applied additional selection criteria. We visually
checked all best-fitting light curves and excluded those targets
that were not well-matched. We did not retain sample objects with
inclinations of less than $50\arcdeg$, since the best-fitting
  parameters for such low inclinations are unreliable.

\begin{deluxetable*}{cccccccccccccc}
\small
\tablecaption{EW-type Parameters \label{tab:Parameters}}
\tablewidth{700pt}
\tabletypesize{\scriptsize}
\tablehead{
\colhead{ID} &
\colhead{$T_1$} &
\colhead{$T_2$} &
\colhead{$q$} &
\colhead{$ M_1$} &
\colhead{$ M_2$} &
\colhead{$ R_1$} &
\colhead{$ R_2$} &
\colhead{$M_{bol1}$} &
\colhead{$M_{bol2}$} &
\colhead{$L_1/(L_1+L_2)$} &
\colhead{$\Omega_1=\Omega_2$} &
\colhead{$A$} &
\colhead{$i$} \\
\colhead{} &
\colhead{$\rm K$} &
\colhead{$\rm K$} &
\colhead{$(\rm M_2/M_1)$} &
\colhead{$M_{\odot}$} &
\colhead{$M_{\odot}$} &
\colhead{$R_{\odot}$} &
\colhead{$R_{\odot}$} &
\colhead{${\rm mag}$} &
\colhead{${\rm mag}$} &
\colhead{} &
\colhead{} &
\colhead{$R_{\odot}$} &
\colhead{$\arcdeg$} 
 }
\startdata
KISOJ000039.63+622214.1& 8157& 6551& 7.5& 0.156& 1.167& 0.494& 1.156& 4.78& 3.894& 0.23& 11.821& 2.005& 57.044\\
KISOJ000048.17+614603.1& 6129& 7699& 0.462& 1.226& 0.567& 1.042& 0.732& 3.668& 3.475& 0.528& 2.79& 2.304& 86.796\\
KISOJ000053.24+613059.8& 4607& 4381& 4.375& 0.251& 1.1& 0.539& 1.041& 5.41& 4.163& 0.248& 8.279& 2.004& 63.157\\
KISOJ000056.84+625228.4& 7337& 7698& 0.3& 1.94& 0.582& 1.929& 1.11& 1.567& 2.567& 0.731& 2.463& 3.946& 56.189\\
KISOJ000123.66+613746.8& 5771& 5477& 0.4& 1.185& 0.474& 1.246& 0.825& 3.822& 4.926& 0.731& 2.645& 2.647& 63.908\\
KISOJ000208.16+630633.5& 5285& 5772& 0.587& 0.963& 0.566& 0.952& 0.746& 4.773& 4.934& 0.548& 3.031& 2.217& 54.22\\
KISOJ000210.22+611924.2& 5312& 5588& 5.0& 0.354& 1.768& 0.894& 1.817& 3.662& 1.991& 0.167& 9.015& 3.405& 58.017\\
KISOJ000233.43+614514.3& 4917& 4658& 4.375& 0.251& 1.098& 0.605& 1.141& 5.332& 4.17& 0.254& 8.147& 2.154& 72.266\\
KISOJ000250.76+624850.0& 5126& 5345& 4.875& 0.213& 1.038& 0.486& 0.977& 6.143& 4.43& 0.174& 8.86& 1.838& 68.057\\
KISOJ000300.42+631913.6& 6261& 5215& 1.95& 0.577& 1.125& 0.806& 1.093& 4.013& 4.062& 0.493& 5.137& 2.455& 83.705\\
KISOJ000308.44+622223.9& 5126& 5016& 7.5& 0.16& 1.201& 0.48& 1.192& 5.66& 3.762& 0.15& 12.146& 2.132& 61.469\\
KISOJ000329.49+613312.2& 5632& 5463& 1.575& 1.01& 1.591& 1.144& 1.396& 2.768& 2.473& 0.425& 4.513& 3.191& 75.15\\
KISOJ000356.37+611452.6& 6356& 5651& 0.237& 2.032& 0.483& 2.089& 1.168& 1.355& 2.785& 0.825& 2.216& 3.837& 67.856\\
KISOJ000359.11+615018.1& 5394& 5659& 0.188& 1.04& 0.195& 1.13& 0.557& 4.42& 5.783& 0.783& 2.141& 2.052& 74.899\\
KISOJ000401.47+613034.7& 4962& 5308& 0.387& 1.115& 0.432& 1.068& 0.692& 4.1& 4.761& 0.65& 2.641& 2.28& 81.237\\
KISOJ000417.46+620836.4& 6733& 6526& 9.5& 0.151& 1.438& 0.549& 1.471& 4.978& 2.938& 0.129& 14.35& 2.52& 66.477\\
KISOJ000429.77+630856.5& 5771& 5784& 3.25& 0.385& 1.25& 0.834& 1.401& 4.76& 3.579& 0.258& 6.796& 2.807& 68.041\\
KISOJ000441.52+623631.7& 6829& 6771& 1.05& 1.838& 1.93& 2.155& 2.197& 1.598& 1.59& 0.496& 3.562& 5.072& 78.901\\
KISOJ000501.27+611509.9& 4729& 4840& 0.125& 1.822& 0.228& 2.158& 0.939& 1.854& 3.63& 0.835& 1.958& 3.634& 71.187\\
KISOJ000505.12+630032.3& 5903& 5052& 1.3& 0.896& 1.165& 0.915& 1.037& 3.545& 3.9& 0.571& 4.277& 2.621& 57.358\\
 \nodata & \nodata & \nodata & \nodata & \nodata & \nodata   & \nodata & \nodata & \nodata &    \nodata & \nodata & \nodata  & \nodata\\
\enddata
\tablecomments{This table is available in its entirety in machine-readable form.}
\end{deluxetable*}

The results from our light-curve modeling are presented in Table
\ref{tab:Parameters}. The first column corresponds to the same order
as that in Table \ref{tab:1}. The input effective temperature of the
primary component, $T_{1o}$, was determined from its color, and the
temperature of the other star, $ T_{2o}$, was obtained from the best
fit. Note that the color index was used to calculate the
temperature of the primary star, which may cause a level of bias in
the temperatures of both components. We therefore calculated the
combined temperature, $T_{\rm
  comb}=[\frac{L_1+L_2}{L_1/T_{1o}^4+L_2/T_{2o}^4}]^{1/4}$, where
$T_{1o},T_{2o},L_1,L_2$ are the temperatures and luminosities of both
components. Next, the temperatures were corrected by dividing
them by $T_{\rm comb}/T_{1o}$. A second run was initiated using these
new temperatures. The results of the second run, $T_1$ and $T_2$, are
the temperatures of the two stars in Table \ref{tab:Parameters}. In
addition, $q$ is the best-fitting mass ratio resulting in the smallest
residuals, and ${(M_1, M_2, R_1, R_2)}$ are the absolute parameters of
both stars in units of solar masses and solar radii, where we
derived the masses from the mass–-luminosity relation
\citep{2013MNRAS.430.2029Y} and the mass ratio $q$. $M_{\rm bol1}$
and $M_{\rm bol2}$ are the absolute bolometric magnitudes of the
primary and secondary stars, respectively. $\Omega_1=\Omega_2$ are the
potentials of both stars. $L_1/(L_1+L_2)$ is the ratio of the bandpass
luminosities of the primary star to the total bandpass luminosity, and
$A$ is the length of the semi-major axis, in solar radii, of the
relative orbit, which is the sum of the two stars' absolute semi-major
axes, $A=a_1+a_2$; $i$, in the final column, is the binary orbital
inclination with respect to the plane of the sky.

We carried out a photometric analysis of all EW-type light curves in
the Galactic plane and derived their parameters. This data set will be
valuable for further analysis and comparisons with other EW-type EBS.

\subsection{Cross-check with {\sl Gaia}}\label{sec:gaia}

The {\sl Gaia} mission
\citep{2016A&A...595A...2G,2018A&A...616A...1G,2020arXiv201201533G}
represents a leap forward for tests of stellar and Galactic
astrophysics. In particular, {\sl Gaia} parallaxes, with
precisions of $30 \mu$as or better for sources with $G\le15$ mag,
can be used to solve the greatest and most challenging problems in
stellar astrophysics. However, there is clear evidence of the
presence of systematic errors in {\sl Gaia} parallaxes.
\citet{2018A&A...616A...2L} found a general, systematic parallax
offset of $\Delta \pi=29$ $\mu \rm as$ from their quasar catalog.
The {\sl Gaia} eDR3 parallax solution is a significant
improvement compared with that affecting {\sl Gaia} Data Release 2
(DR2): a typical 20\% parallax improvement has been reported for
{\sl Gaia} eDR3 quasars \citep{2020arXiv201206242F}. However,
despite being the best benchmark for {\sl Gaia} parallaxes, quasars
(typically with $G>17$ mag) are rather faint and their color
distribution does not match well the stellar color
distribution. Clearly, independent assessment of {\sl Gaia} parallaxes
is important to fully characterize any lingering systematic
errors. Since EW-type EBS are among the most numerous variables in the
Milky Way that have independently determined distance measurements in
the solar neighborhood, here we will use our EW-type distances to
check for zero-point offsets in the {\sl Gaia} eDR3 parallaxes.

\begin{figure}[htbp]
\centering
\begin{minipage}[t]{0.48\textwidth}
\centering
\includegraphics[width=1\textwidth]{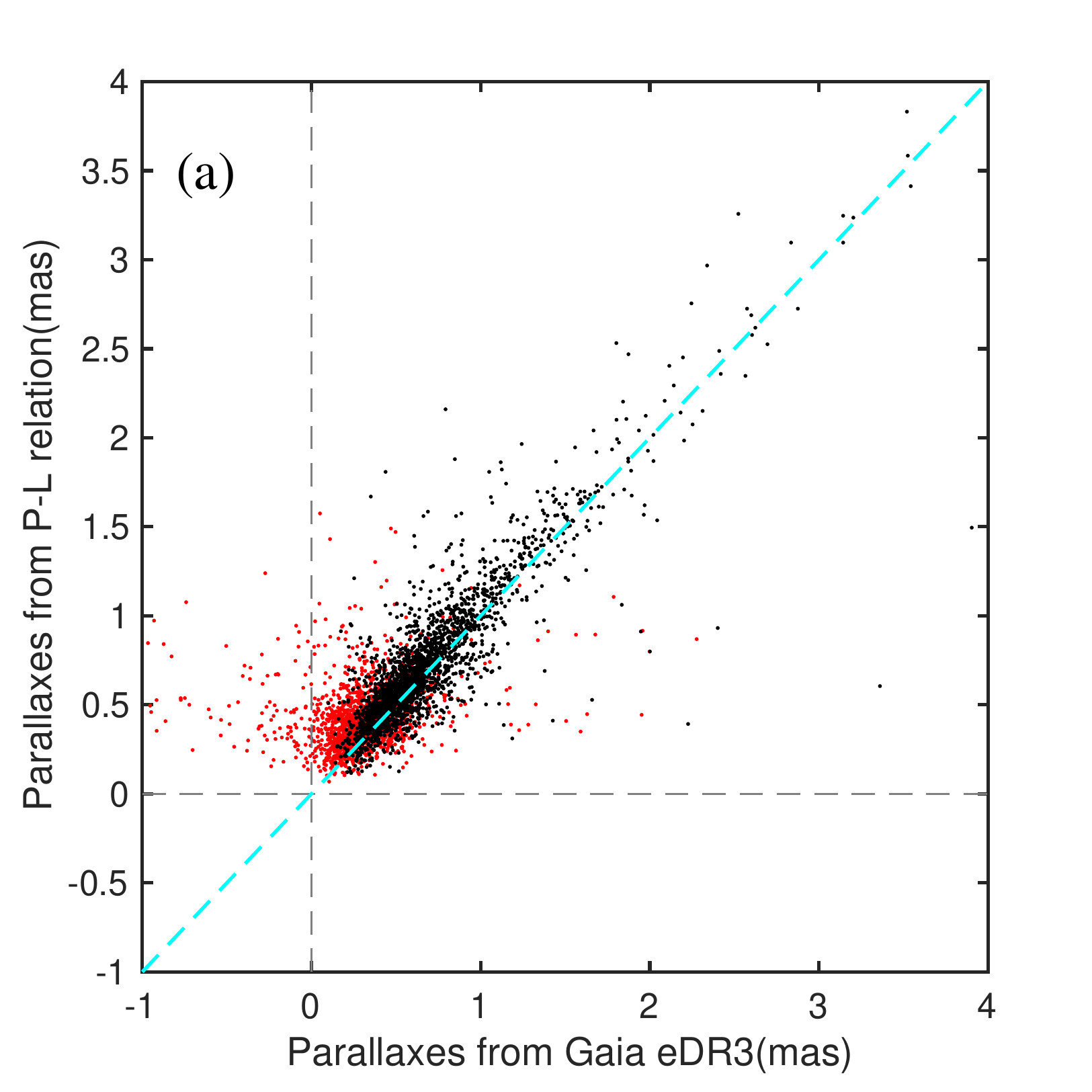}
\end{minipage}
\begin{minipage}[t]{0.48\textwidth}
\centering
\includegraphics[width=1\textwidth]{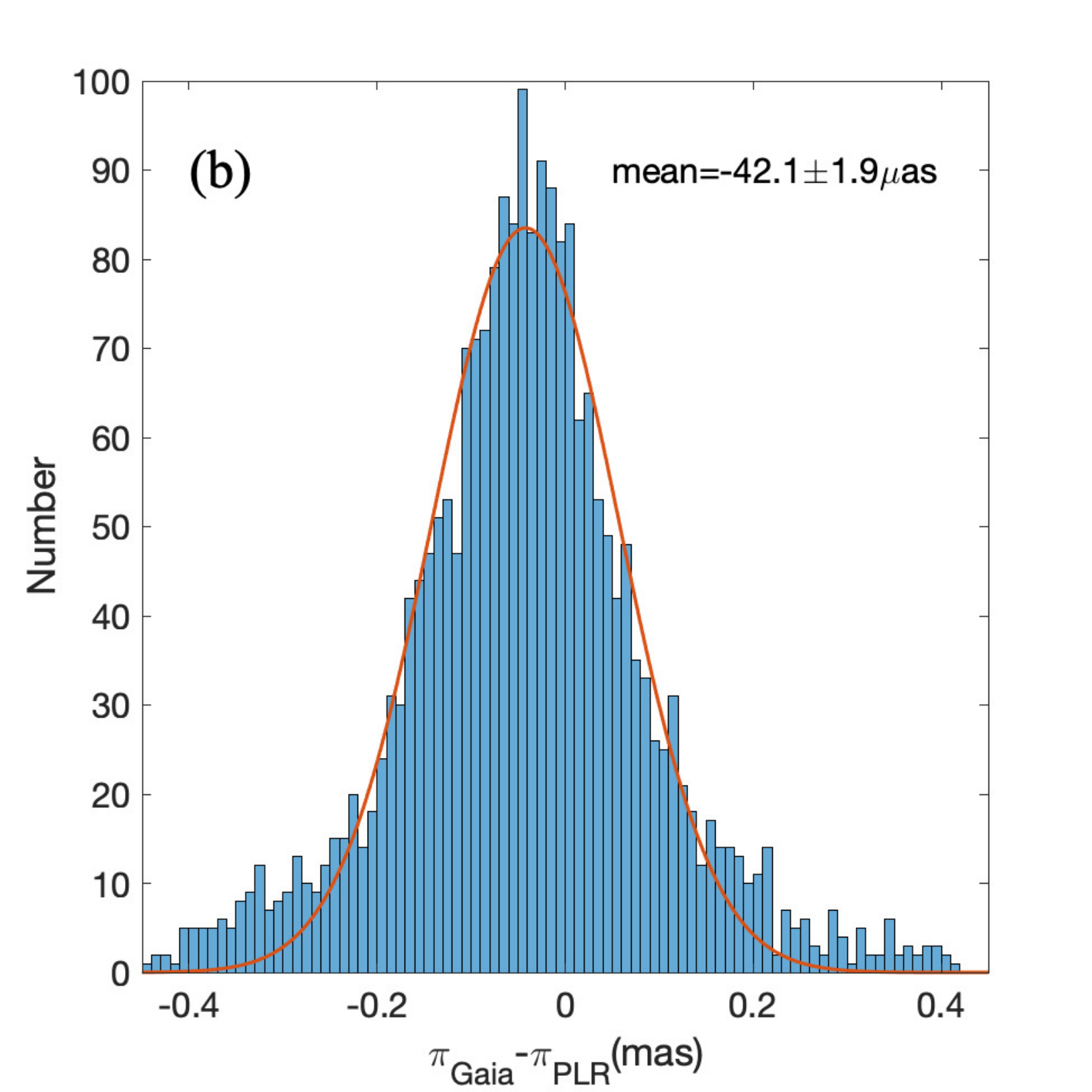}
\end{minipage}
\begin{minipage}[t]{0.48\textwidth}
\centering
\includegraphics[width=1\textwidth]{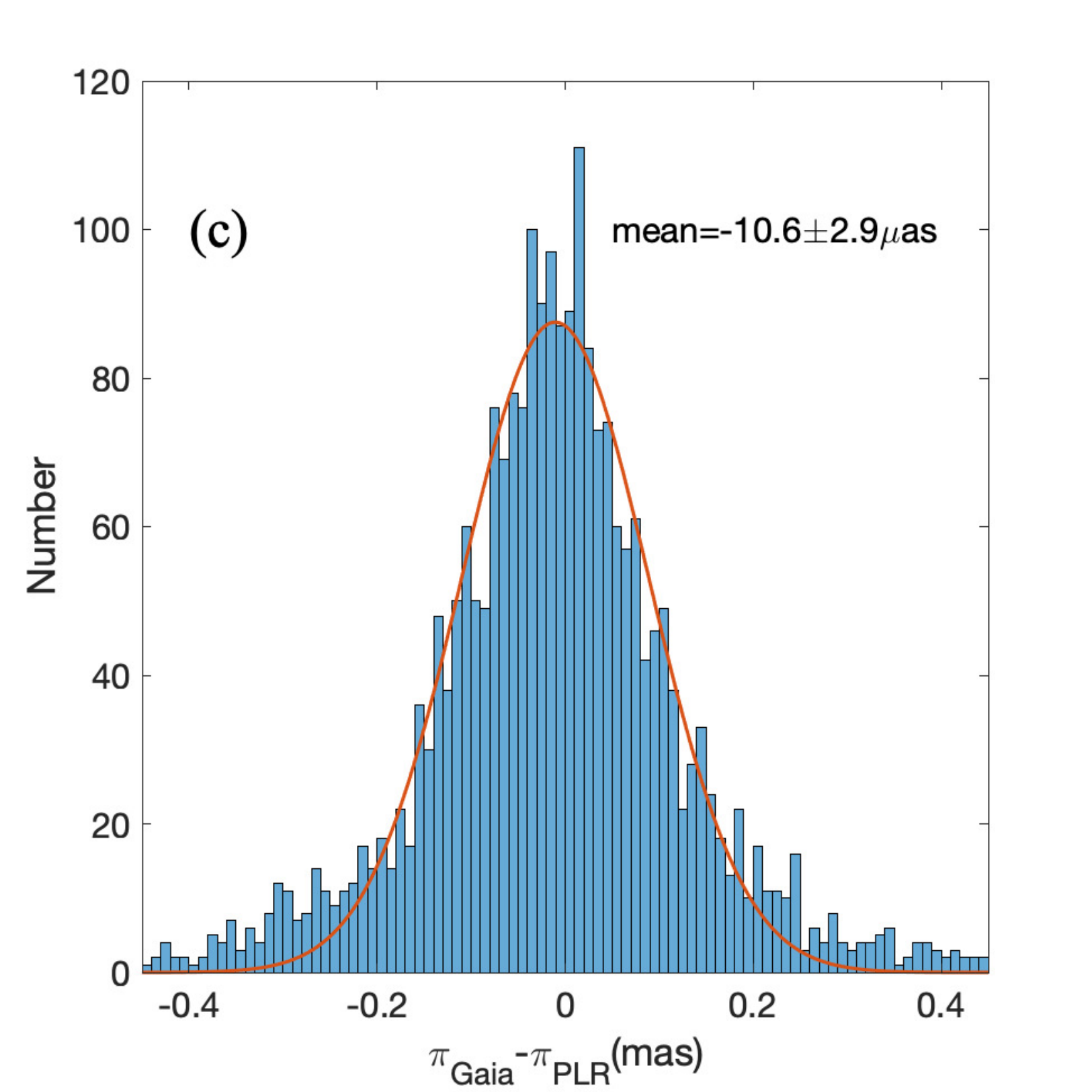}
\end{minipage}
\begin{minipage}[t]{0.48\textwidth}
\centering
\includegraphics[width=1\textwidth]{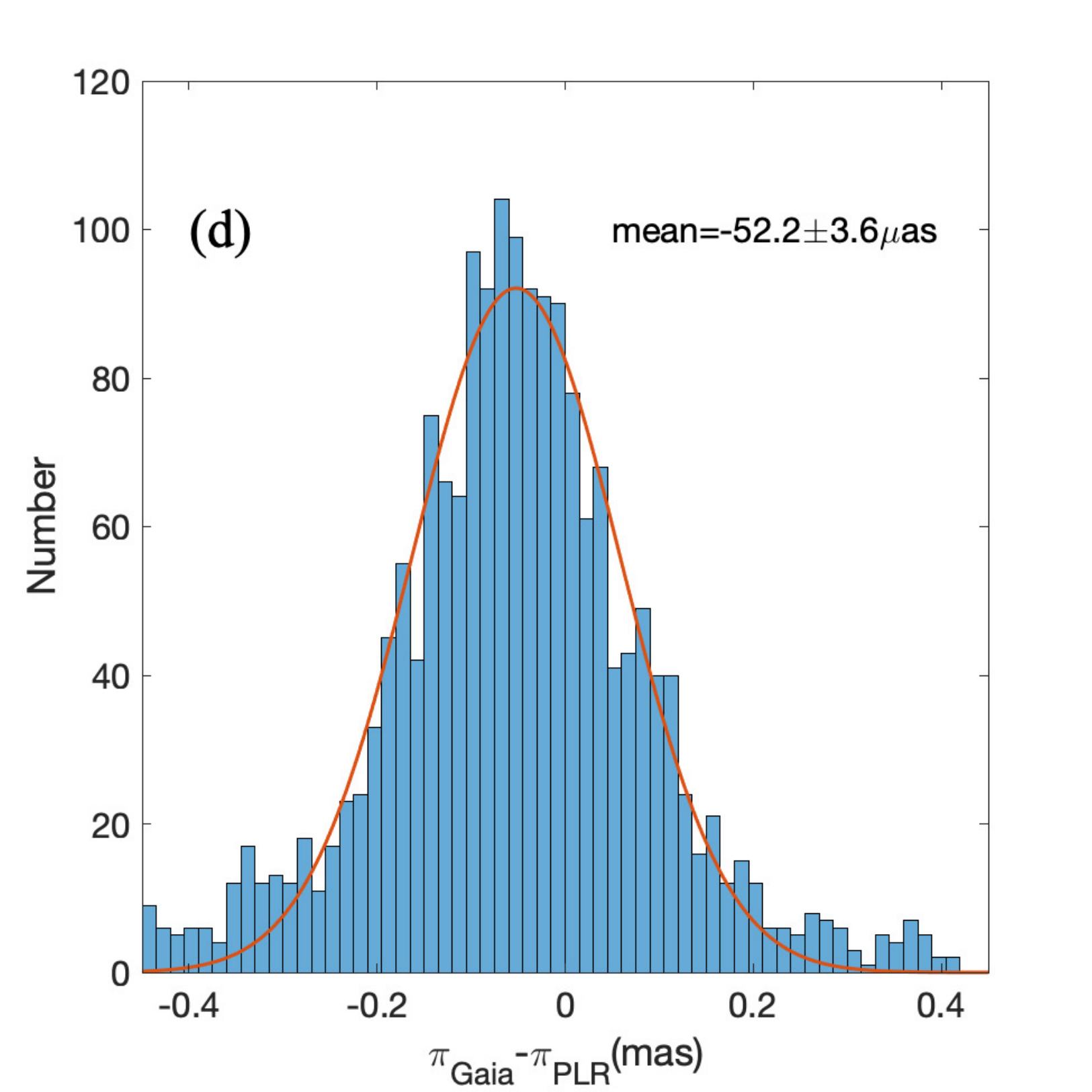}
\end{minipage}
\caption{Comparison of predicted EBS parallaxes derived from the PLR
versus {\sl Gaia} eDR3 parallaxes. (a) Direct comparison
for all EW types. Red points have uncertainties exceeding 20\% or
negative parallaxes. The cyan dashed line is the one-to-one locus.
(b) Distribution of the parallax offsets affecting {\sl Gaia}
eDR3, $\Delta \pi({\rm Gaia}_{\rm eDR3}-{\rm PLR})=-42.1\pm1.9\mu$as. 
(c) Parallax offsets of corrected {\sl
  Gaia} eDR3 parallaxes \citep{2020arXiv201201742L}, $\Delta
\pi({\rm Gaia}_{\rm eDR3}^{\rm corr}-{\rm PLR})=-10.6\pm2.9\mu$as. 
(d) Equivalent distribution of {\sl Gaia}
DR2 parallaxes, $\Delta \pi({\rm Gaia}_{\rm DR2}-{\rm PLR})=-52.2\pm3.6\mu$as. 
The red curves in panels (b), (c),
and (d) represent the best-fitting Gaussian
distributions.\label{fig:8}}
\end{figure}

Of the 3996 EW-type EBS with distances from the PLR, 3920 have
parallaxes available in {\sl Gaia} eDR3. Figure \ref{fig:8}(a)
shows a direct comparison of the EBS parallaxes derived from the PLR
versus {\sl Gaia} eDR3 parallaxes. The cyan dashed one-to-one
line is meant to clearly show the extent of the offset in
distances. The deviation between the samples hints at a clear offset
in parallaxes, in the sense that the PLR parallaxes are larger on
average than their {\sl Gaia} counterparts. After excluding objects
with large errors and negative parallaxes ($\sigma_\pi/\pi>0.2$ and
$\pi<0$), marked as red points in Figure \ref{fig:8}(a), 2334
EW-type EBS were left. Figure \ref{fig:8}(b) presents the distribution
of parallax differences, $\pi_{\rm Gaia}-\pi_{\rm PLR}$. The
distribution appears to be a roughly symmetric, normally distributed
offset in the negative direction. The mean offset is $\Delta
\pi=-42.1\pm1.9\mu$as, where the error is the standard error on the
mean for all sample objects. In other words, the {\sl Gaia} parallaxes
are systematically smaller.

The {\sl Gaia} team has released a model allowing us to adjust this
zero-point offset, which was based on an analysis of quasars, binary
stars, and stars in the Large Magellanic Cloud
\citep{2020arXiv201201742L}. We also compared the parallaxes implied
by our EW-type distances with the corrected {\sl Gaia} parallaxes (see
Figure \ref{fig:8}c) and found an offset of $\Delta
\pi=-10.9\pm2.9\mu$as. This suggests that the parallax zero-point
correction provided by the {\sl Gaia} team significantly reduces but
may not fully eliminate the prevailing bias in the {\sl Gaia} eDR3
parallaxes.

Quantification of the parallax systematics in {\sl Gaia} DR2 was
based on careful analysis of a series of tracer objects, yielding:
Cepheids, $\Delta \pi=-46\pm13\mu$as \citep{2018ApJ...861..126R};
nearby bright EBS,$\Delta \pi= -82\pm33\mu$as
\citep{2018ApJ...862...61S}; stars with asteroseismically determined
radii, $\Delta \pi=-52.8\pm2.4\mu \rm as \, (statistical)\pm8.6\mu
\rm as \, (systematic)$ \citep{2019ApJ...878..136Z}; and Bayesian
distances for a radial velocity sample, $\Delta \pi= -54\pm6.0\mu$as
\citep{2019MNRAS.487.3568S}). For comparison, we similarly used our
EBS sample as parallax tracer and found $\Delta
\pi=-52.2\pm3.6\mu$as. This offset is similar to those of other
authors' concurrent, independent analyses based on different
benchmark samples
\citep[e.g.,][]{2019ApJ...878..136Z,2019MNRAS.487.3568S}. Our result
should have similar systematic uncertainties as those determined by
\citet{2019ApJ...878..136Z} and \citet{2019MNRAS.487.3568S}, since
the maximum parallax differences among these three tracer
populations is less than 2$\mu$as.

We will now estimate the likely systematic error range pertaining to
our PLR-based EW-type distances. Five fundamental issues affect the
level of the systematic uncertainties, including (i) the offset in
the PLR, (ii) its internal spread, (iii) uncertainties in the
photometric zero points, (iv) errors in our extinction evaluation,
and (v) third-component effects. 

The PLR was obtained from 183 objects within 330 pc with an average
parallax of 5.46 mas. Considering a 30 $\mu$as systematic
uncertainty \citep{2016A&A...595A...2G}, the systematic error
contributed by the offset is about 0.5\%. The systematic error
associated with the internal spread in the PLR is of order
$0.21/\sqrt{183}\approx0.016$ mag. Here, 0.21 mag is the mean spread
in the $I J H K_{\rm s}$-band PLRs. As regards the photometric zero
points, the $I$-band systematic error is most important, given that
the $J H K_{\rm s}$-band PLR and KISOGP EW data both came from
2MASS; our $I$-band photometric data for all EBS PLRs was based on
the USNO-B catalog. We correlated our $I$-band data for all sample
EBS with that published in the USNO-B catalog. We found that the
KISOGP $I$-band magnitudes are systematically 0.009 mag brighter
than the corresponding USNO-B $I$-band magnitudes for the 6894
targets in common.

Since we make use of the $I J H K_{\rm s}$ bands, here we consider
an average extinction corresponding to $0.225 A_V$. We also adopt a
10\% uncertainty in the extinction determination, which reflects
uncertainties owing to the choice of extinction law. For a mean
observed extinction of $A_V = 1.94 mag$, these choices lead to an
uncertainty of 1.94 mag $\times$ 0.225 $\times$ 10\% = 0.0437
mag. Next, if a third component can be discerned, a study of 75
nearby EW-type EBS \citep{2006AJ....132..650D} has shown that this
will affect the systematic uncertainty at the median distance of our
sample by about 0.3\%. As neither our KISOGP sample objects nor
those contributing to the PLR explicitly excluded third-body
effects, the bias caused by third- or higher-order multiplicity
should not differ much between both sets of EBS. We estimate that
the associated systematic uncertainty is less than 0.3\%. As such,
the systematic uncertainty range affecting our results is
$\sigma=581\mu {\rm as} \times [(0.005)^2+ (0.016 \times
\ln(10)/5)^2+(0.009 \times \ln(10)/5)^2+(0.0437 \times
\ln(10)/5)^2+(0.003)^2]^{1/2}=12.9\mu$as.

The offset we found based on our catalog of 2334 EW-type EBS is well
within the prevailing uncertainties and also fully consistent with a
systematic error below $100\mu$as as reported by the {\sl Gaia}
team. Our EW-type EBS in the Galactic plane result in a larger
offset than that derived from quasars ($-17 \mu$as), but the offset
can be reduced significantly by application of the official parallax
zero-point correction. The {\sl Gaia} team found that the parallax
zero-point depends on a target's magnitude, color, and position, and
hence the small difference we found between EBS and quasars is not
surprising. The difference may come from the details of the
distribution of the parallax zero-point, since quasars are fainter
and bluer than our EBS, which are all located in the Galactic plane.
In addition, for faint sources outside the Galactic disk parallax
calibrations are estimated directly from the quasar sample. However,
the parallax bias for objects in the Milky Way is derived
indirectly, based on binary stars and stars in the Large Magellanic
Cloud.

\citet{2021arXiv210103425S} found a mean parallax offset of
$-37\pm20\mu$as, which decreased to $-15\pm18\mu$as following the
official corrections based on 76 EBS. Their result matches our
result well, which is particularly encouraging since our result is
affected by small statistical errors and based on a large sample.
Our EBS sample can be used as useful tracers for further work on the
zero-point offset of {\sl Gaia} parallaxes. In addition, our result
is based on EW-type EBS in the Galactic plane, which most other
studies try to avoid. It can therefore serve as a useful reference
for other studies that need to deal with the {\sl Gaia} zero-point
offset in the Galactic plane and complement studies of the
zero-point distribution across the sky.

In conclusion, we found a {\sl Gaia} eDR3 zero-point offset of 
$\Delta \pi=-42.1\pm1.9\mbox{ (stat.)}\pm12.9\mbox{(syst.)} \mu$as, 
based on 2334 EW-type EBS covering a wide range of
magnitudes and extinction values in the northern Galactic plane. The
official parallax zero-point correction can significantly reduce the
bias in eDR3 parallaxes to $ -10.9\pm2.9\mbox{(stat.)}\pm12.9
\mbox{ (syst.)}\mu$as (for our sample).

\subsection{Extinction compared with 3D extinction map} 

\begin{figure}
\centering
\includegraphics[width=10cm]{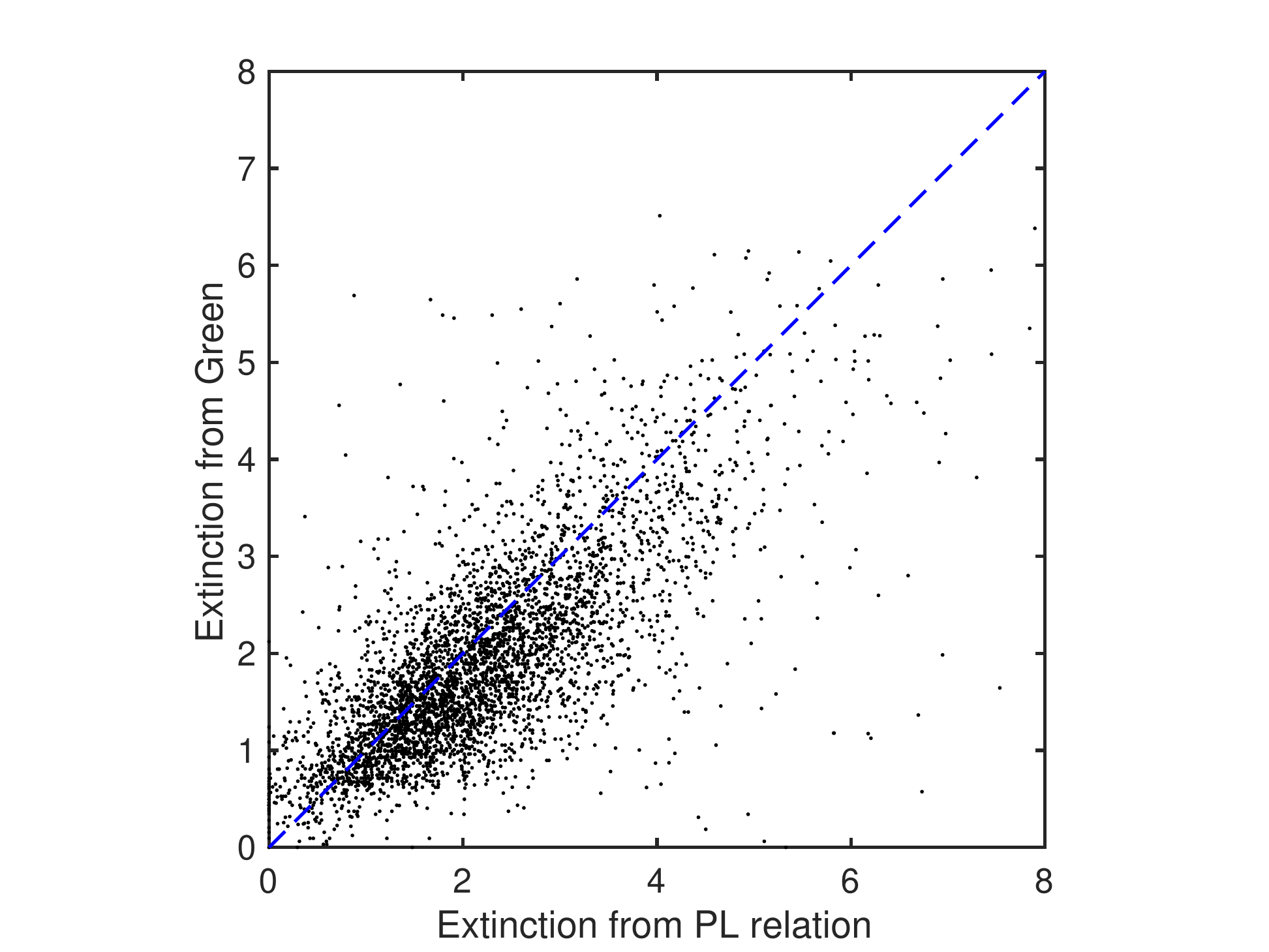}
\caption{Extinction comparison ($A_V$, mag) between the values derived
  from the PLR and from \citet{2019ApJ...887...93G}. \label{fig:11}}
\end{figure}

\citet{2019ApJ...887...93G} presented a three-dimensional (3D) dust
reddening map of the northern sky derived from {\sl Gaia} parallaxes
and stellar photometry from Pan-STARRS 1 and 2MASS. Green et al.'s 3D
extinction map happens to overlap with the KISOGP survey's spatial
coverage. In Figure \ref{fig:11}, we compare the extinction derived
from our PLR analysis with that from the 3D extinction map.

As shown in Figure \ref{fig:11}, a linear correlation is clearly
discernible. The extinction values from both studies fit well within
the relevant scatter envelopes. This scatter may originate from
various sources. First, the error in the extinction from both our PLR
analysis and that in the 3D extinction map cannot be ignored. Second,
the error in distance used as input into the 3D extinction map may
exacerbate the intrinsic errors in the extinction from the 3D
map. Finally, the distances and spatial resolutions pertaining to the
3D map are coarse, with distance modulus steps of 0.5 mag and angular
resolutions of several arcminutes. These steps lead to non-negligible
errors when comparing the extinction values for single objects,
particularly for our sample in the Galactic plane where the extinction
is usually high. In any case, given the prevailing uncertainties, both
methods are mutually consistent.

\section{Conclusion} \label{sec:conclusion}

We have presented a new catalog of EBS in the northern Galactic plane
based on the KISOGP survey. We visually identified 7055 EBS spread
across $\sim$330 deg$^2$, including 4197 EW-, 1458 EB-, and 1400
EA-type EBS. For all sample objects, we used their $I$-band light
curves to determine accurate parameters, including their periods
(accurate to better than the fifth decimal place), a reference $t_0$,
the depths of the two eclipses, etc.

We also examined the spatial distribution of our sample objects. We
found an inhomogeneous density distribution of EBS in the thin disk at
different Galactic longitudes. In addition, we found a random
distribution of the ratios of the eclipse depths for EA types and a
concentration tending to unity for EW-type EBS. Moreover, we checked
that the distributions of their periods vary among different EBS
types, increasing from EW to EB and EA type. We also obtained the
distribution of the eclipse depths, the $I$-band magnitudes,
etc. Finally, we tested the level of contamination of our sample by
other types of variables, which we found to be negligible.

We derived the distances and extinction values pertaining to the EW
types in our sample using their PLRs, reaching distances in excess of
6 kpc and $V$-band extinction values exceeding $A_V = 9$ mag. We
combined our EBS sample with HMSFRs and Cepheids to trace the
structure of the thin disk.  Stars of the same type (including but
not limited to EBS) tend to cluster on spatial scales of several
hundred parsecs. Using different tracers, we revealed some
structural properties of the thin disk.

As an independent distance measurement, our EBS distance analysis
offers a complementary measurement of the global parallax offset
affecting {\sl Gaia} eDR3. We found $\Delta \pi=-42.1\pm1.9\mbox{
(stat.)}\pm12.9\mbox{ (syst.) } \mu$as based on a careful
analysis of 2334 EW-type EBS. Our newly derived offset is consistent
with the results from the {\sl Gaia} team. We also found that the
official parallax zero-point correction can significantly reduce the
bias affecting the eDR3 parallaxes.  Finally, we performed a
photometric analysis of all EW-type light curves using the W--D method
to derive individual system parameters, and we cross checked our
extinction values with Green et al.'s 3D extinction map.

\acknowledgments

We acknowledge long-term support for the KISOGP project from the staff
at Kiso Observatory, Japan. We are grateful for research support from
the National Key Research and Development Program of China through
grants 2019YFA0405500 and 2017YFA0402702. This work was also partially
supported by the National Natural Science Foundation of China (NSFC)
through grant 11973001. N. M. acknowledges financial support from
Grants-in-Aid (Nos. 26287028 and 18H01248) from the Japan Society for
the Promotion of Science. X. C. acknowledges support from NSFC grant
11903045.
\vspace{5mm}

\bibliographystyle{aasjournal_V1}
\bibliography{KISOGP_EBs}

\end{document}